\documentclass[11pt,a4paper]{article}
\usepackage{jheppub}
\usepackage{physics}
\usepackage{amsmath}
\usepackage{grffile} %Needed for using file with multi-dot filename
\usepackage{yfonts}
\usepackage[title]{appendix}

\title{Nambu-Goldstone modes in non-equilibrium systems from AdS/CFT correspondence}
                                           \author[a]{Shuta Ishigaki}
                                           \author[b]{and Masataka Matsumoto}
                                           \affiliation[a]{Department of Physics, Chuo University,\\
                                           1-13-27 Kasuga, Bunkyo-ku,
                                           Tokyo 112-8551, Japan}
                                           \affiliation[b]{Department of Mathematics, Shanghai University,\\
                                           Shanghai 200444, China}
                                           \emailAdd{ishigaki@phys.chuo-u.ac.jp, matsumoto@shu.edu.cn}
                                           \abstract{
                                           We investigate dispersion relation of Nambu-Goldstone modes in a dissipative system realized by the AdS/CFT correspondence. We employ the D3/D7 model which represents ${\cal N}=4$ supersymmetric Yang-Mills theory coupled to ${\cal N}=2$ flavor fields. If we consider massless quarks in the presence of an external magnetic field, the system exhibits the phase transition associated with the spontaneous symmetry breaking of the chiral symmetry. We find that the Nambu-Goldstone modes show a diffusive behavior in the dispersion relation, which agrees with that found with the effective field theory approach. We also study a non-equilibrium steady state which has a constant current flow in the presence of an external electric field. In a non-equilibrium steady state, we find that the Nambu-Goldstone modes show a linear dispersion in the real part of the frequency in addition to the diffusive behavior. Moreover, we analyze the linear dispersion of the Nambu-Goldstone modes in the hydrodynamic approximation. As a result, we find that the linear dispersion can be written as the analytic functions of quantities in the dual field theory. Our results imply that such a linear dispersion is a characteristic behavior of Nambu-Goldstone modes in a non-equilibrium steady state.
                                           }
                                           \arxivnumber{}
                                           
\begin{document}
                                           \maketitle
                                           \flushbottom
\section{Introduction}
The AdS/CFT correspondence is a powerful tool to study the strongly coupled gauge theory\,\cite{Maldacena1997,Gubser1998,Witten1998}. This duality has been applied to various areas in physics such as condensed matter physics or QCD (for example, see Refs.\,\cite{CasalderreySolana2011,Ammon2015,Zaanen2015,Hartnoll2016}). Particularly, one of the most significant benefits is that one can analyze a system far from equilibrium\,(for example, see Refs.\,\cite{Hubeny2010,Liu2018,Kundu2019}). The comprehensive understanding of a far-from-equilibrium system, such as a universal property which does not depend on the details of the non-equilibrium system, is still lacking even in a non-equilibrium steady state\,(NESS) system. A NESS system carries various currents, for instance an electric current and/or a heat current, but has no time-dependent observables.

Moreover, the AdS/CFT correspondence provides us a prescription for the calculation of the retarded Green's function at finite temperatures\,\cite{Son2002}. According to this prescription, one can calculate several transport coefficients such as the shear viscosity or the conductivity. Also, the dispersion relation for a small fluctuation at finite temperature can be obtained from quasinormal frequancies. In the gravity theory, quasinormal modes\,(QNMs) are defined by solutions to linearized equations for fluctuations in a black hole spacetime with specific boundary conditions imposed. It was pointed out that QNMs with specific boundary conditions correspond to poles of the retarded Green's function in a holographically dual theory\,\cite{Son2002}.

As mentioned above, various systems, including NESS systems, can be realized in the context of the AdS/CFT correspondence. An important example is a system which exhibits the phase transition associated with the chiral symmetry breaking in the dual field theory. For instance, the D3/D7 model represents the large $N_{c}$ ${\cal N}=4$ gauge theory with quenched ${\cal N}=2$ quark matter\,\cite{Karch2002}. In this model, the chiral symmetry is broken in the system with a finite chemical potential in the presence of an external magnetic field at finite temperature\,\cite{Evans2010}. Furthermore, this chiral symmetry breaking is also realized in a NESS system which has a constant current flow in the presence of an external electric field\,\cite{Evans2011mu}.

In this paper, we address the issue of Nambu-Goldstone\,(NG) modes associated with this chiral symmetry breaking.\footnote{In\,\cite{Amado2009}, the dispersion relation of NG mode in a dissipative system was studied by using the holographic superconductor model in which the $U(1)$ global symmetry of the dual system is broken. In our study, we focus on a spontaneous chiral $U(1)$ symmetry breaking in the D3/D7 model.} Particularly, we investigate the dispersion relation of the NG modes by analyzing the corresponding QNMs in the dual gravity theory since it has not been explored in the gravity side. If we consider a fluctuation of the background in the gravity picture, it is dissipated into the black hole while the background is stationary. For this reason, it is expected that the dispersion relation of NG modes corresponds to that in a dissipative system. In recent studies based on the effective field theory approach in the field theory side\,\cite{Minami2018,Hongo:2019qhi}, a spontaneous symmetry breaking of internal symmetry and NG modes in a dissipative system were studied by using simple models. It was shown that the NG mode represents different dispersion relations depending on whether the Noether charge is conserved or not. We will show that our results agree with them.

%In this paper, we study the feature of this NG mode by analyzing the quasi-normal modes (QNMs). In the gauge/gravity duality, the QNMs correspond to the poles of the Green's functions computed by following the holographic prescription\,\cite{Son2002}. Computing several lower QNMs, we study the dispersion relation of the NG mode in our dissipative system.

This paper is organized as follows, In section \ref{sec2}, we briefly review the system in which the chiral symmetry is spontaneously broken by applying an external magnetic field in the presence of a finite charge density. This system is in equilibrium. In section \ref{sec3}, we study the dispersion relation of NG mode by considering fluctuations of a pseudo-scalar field and a gauge field in the equilibrium background. In section \ref{sec4}, we consider the NESS background by applying an external electric field. We also study the dispersion relation of NG mode in the NESS system. The section \ref{sec5} is devoted to conclusion and discussion.

\section{Holographic setup} \label{sec2}
In our study, we employ the D3/D7 model. The system consists of a stack of $N_{c}$ D3-branes and $N_{f}$ D7-branes. If we take the limit of $N_{f} \ll N_{c}$ and large 't\,Hooft coupling $\lambda=g_{\rm YM}^{2}N_{c}$ with $N_{f}$ fixed, the D7-branes can be regarded as probes in the black D3-brane geometry. The dual field theory of the D3/D7 model is ${\cal N}=4$ $SU(N_{c})$ supersymmetric Yang-Mills theory coupled to flavor fields preserving ${\cal N}=2$ supersymmetry. In this section, we briefly review the spontaneous symmetry breaking of the $U(1)$ chiral symmetry induced by the magnetic field in the D3/D7 model.

\subsection{Background}
 We assume that the bulk geometry is given by a five-dimensional AdS-Schwarzschild black hole\,(AdS-BH) and an $S^{5}$:
\begin{equation}
		ds^{2}= \frac{1}{u^{2}}\left(-f(u)dt^{2}+d\vec{x}^{2}+\frac{du^{2}}{f(u)}\right)+d\Omega_{5}^{2},
	\label{eq:metric}
\end{equation}
where
\begin{equation}
	f(u)=1-\frac{u^{4}}{u_{H}^{4}}.
\end{equation}
Here, $(t,\vec{x})=(t,x,y,z)$ is the coordinate of the dual gauge theory in the (3+1)-dimensional spacetime and $u$ is the radial coordinate. The black hole horizon is located at $u=u_{H}$ and the AdS boundary is located at $u=0$. The Hawking temperature is given by $T=1/(\pi u_{H})$, which corresponds to the heat bath temperature in the dual field theory. The metric of $S^{5}$ part is written by
\begin{equation}
	d\Omega_{5}^{2}=d\theta^{2}+\sin^{2}\theta d\psi^{2}+\cos^{2}\theta d\Omega_{3}^{2},
\end{equation}
where $d\Omega_{3}$ is the line element of the $S^{3}$ part. For simplicity, we set the radius of $S^{5}$ part to be 1. In this study, we introduce a single D7-brane\,($N_{f}=1$) which fills the AdS$_{5}$ part and wraps the $S^{3}$ part of the $S^{5}$. The configuration of the D7-brane is determined by $\theta$ and $\psi$.

The D7-brane action is given by the Dirac-Born-Infeld\,(DBI) action:
\begin{equation}
	S_{D7}=-T_{D7} \int d^{8}\sigma \sqrt{-\det \left(g_{ab}+(2\pi\alpha')F_{ab} \right)}
	+ \frac{(2\pi\alpha')^2}{2}T_{\mathrm{D7}} \int C^{(4)}\wedge F \wedge F,
\end{equation}
where $T_{D7}$ is the D7-brane tension given by $T_{D7}=(2\pi)^{-7}(\alpha')^{-4}g_{s}^{-1}$. $g_{ab}$ is the induced metric of the D7-brane:
\begin{equation}
	g_{ab}=\frac{\partial X^{\mu}}{\partial \sigma^{a}}\frac{\partial X^{\nu}}{\partial \sigma^{b}} G_{\mu\nu},
\end{equation}
where $\sigma^{a}$ denotes the worldvolume coordinates on the D7-brane with  $a,b=0,\cdots,7$ and $X^{\mu}$ denotes the target-space coordinates with $\mu,\nu=0,\cdots,9$. $G_{\mu\nu}$ is the background metric given by Eq.\,(\ref{eq:metric}). $F_{ab}$ is the field strength of the $U(1)$ gauge field on the D7-brane.
\(C^{(4)}\) is the background Ramond-Ramond 4-form potential which is written as
\begin{equation}
	C^{(4)} =
	r^4 dt \wedge dx \wedge dy \wedge dz
	+ \cos^4\theta d\psi \wedge \epsilon(\mathrm{S}^3),
\end{equation}
where \(\epsilon(\mathrm{S}^3)\) is the volume form of the 3-sphere with unit radius.
Note that we have not taken the Wess-Zumino term into account since it does not contribute to our background setup. In later sections, however, we will consider the contribution of the Wess-Zumino term to the perturbation analysis.
For convenience, we use the static gauge $X^{a}=\sigma^{a}$ and $2\pi\alpha'=1$.

We consider the following ansatz for fields:
\begin{equation}
	\theta=\theta(u), \hspace{1em} \psi=0, \hspace{1em} A_{t}=a_{t}(u), \hspace{1em} A_{y}=Bx.
	\label{eq:ansatz}
\end{equation}
Here, we refer to the scalar field $\theta(u)$ as the {\it embedding function}. Since the D7-brane action is independent of $a_{t}(u)$, there is a conserved quantity $D \equiv \delta S / \delta F_{ut}$ which is corresponding to the charge density in the field theory side\,\cite{Karch2007}. For convenience, we eliminate the variable of $a_{t}(u)$ at the level of the action by performing the Legendre transformation:
\begin{equation}
	\tilde{S}_{D7} = S_{D7} - \int d^{8}\sigma F_{ut} \frac{\delta S}{\delta F_{ut}} = -T_{D7}\int d^{8}\sigma \sqrt{-g_{tt}g_{uu}} \sqrt{D^{2}+g_{xx}(B^{2}+g_{xx}^{2})\cos^{6}\theta}.
	\label{eq:legaction}
\end{equation}
%where $g=\det g_{ab}$.
Then, we consider the equation of motion only for the embedding function $\theta(u)$ obtained from (\ref{eq:legaction}). If we consider the asymptotic behavior of $\theta(u)$ near the AdS boundary, it can be expanded as
\begin{equation}
	\theta(u) = m_{q} u + \theta_{2} u^{3} + \cdots.
	\label{eq:theta}
\end{equation}
Here, according to the AdS/CFT dictionary, $m_{q}$ corresponds to the quark mass and $\theta_{2}$ corresponds to the chiral condensate via following relation\,\cite{Mateos2007}\,(see also \cite{Karch2007}):
\begin{equation}
	\expval{\bar{q}q}={\cal N} \left( 2 \theta_{2}-\frac{m_{q}^{3}}{3} \right),
	\label{eq:qq}
\end{equation}
where
\begin{equation}
	{\cal N}=T_{D7}(2\pi^{2})=\frac{N_{c}}{(2\pi)^{2}}.
\end{equation}
Here, we used the relation of $4\pi g_{s}N_{c} \alpha'^{2}=1$. Hereafter, we set ${\cal N} =1$ for simplicity.
Depending on the boundary condition at the black hole horizon for the bulk part, there are two different possible solutions for the embedding function $\theta (u)$:\,{\it Minkowski embedding} and {\it black hole embedding}. For Minkowski embedding, the D7-brane does not reach the black hole horizon. In order to avoid the conical singularity, we impose the boundary condition $\theta'(u_{\rm max})=0$, where $u_{\rm max}$ is the position at which $\theta ( u_{\rm max}) = \pi /2$ and $u_{\rm max} < u_{H}$. For black hole embedding, on the other hand, the D7-brane reaches the black hole horizon. We impose the boundary condition $\theta'(u_{H})=0$. In terms of the dual field theory, Minkowski embedding and black hole embedding correspond to the meson bound state and the meson melting state, respectively\,\cite{Kruczenski2003}.

%Under these boundary conditions, we numerically solve the equations of motion for $\theta(u)$ by using the shooting method. After we solve them for several values of $B$ and $D$, we pick up solutions in which the quark mass becomes zero.

\subsection{Chiral symmetry breaking}
\begin{figure}[htbp]
	\centering
	\includegraphics[width=12cm]{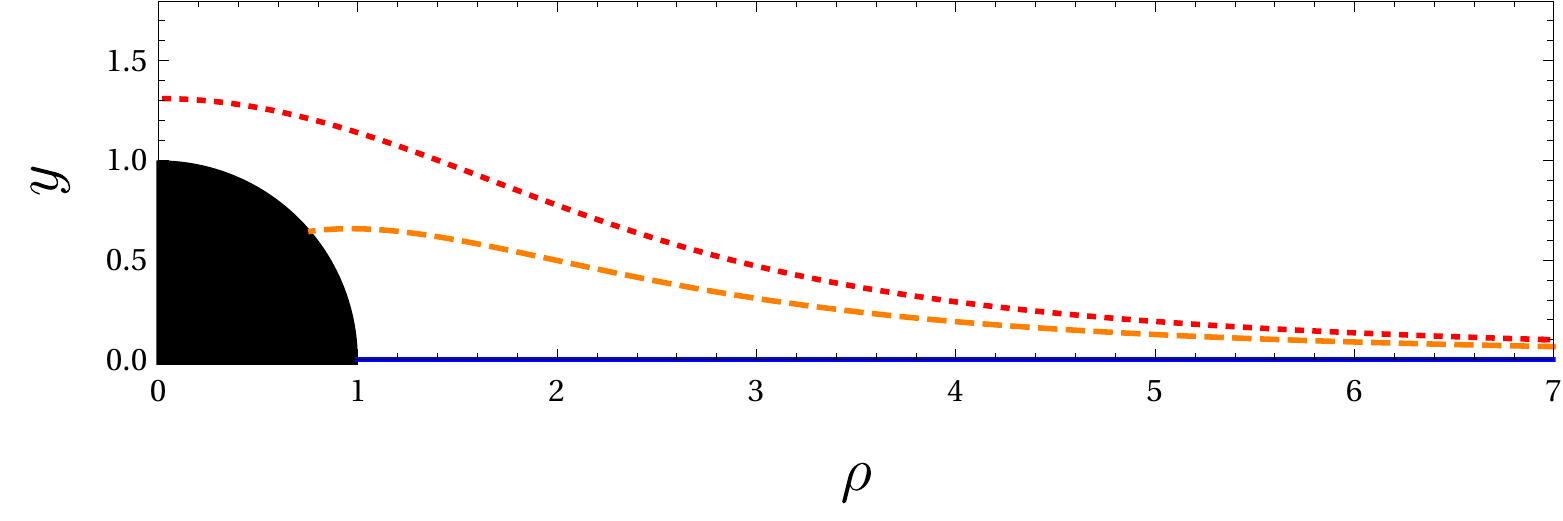}
	\caption{The configurations of the D7-brane which have the asymptotic behavior of $m_{q}=0$. There are two different solutions which are classified by the boundary conditions in bulk:\,Minkowski embedding\,(dotted) and black hole embedding\,(dashed and solid).}
	\label{fig:config}
\end{figure}

In our study, we numerically solve the equation of motion for $\theta(u)$ by shooting method from $u_{\rm max}$ or $u_{H}$. To do so, we set $u_{\rm max}$ to some value for Minkowski embedding or $\theta(u_{H})=\theta_{0}$ for black hole embedding in addition to the above conditions. After we obtain the numerical solutions with various values of $B$, $T$ and $D$, we calculate the quark mass $m_{q}$ and the chiral condensate $\expval{\bar{q}q}$ from Eqs.\,(\ref{eq:theta}) and (\ref{eq:qq}). In this paper, we focus on the solutions which correspond to the massless quark in the dual field theory. We show the typical configurations of the D7-brane in Fig.\,\ref{fig:config}.
Here, we employ the coordinates of $(\rho,y,w)$ instead of $(u,\theta,\psi)$:
\begin{equation}
	\rho=\frac{\cos \theta}{u}, \hspace{1em} y=\frac{\sin\theta\cos\psi}{u}, \hspace{1em} w=\frac{\sin\theta\sin\psi}{u}.
	\label{eq:coord}
\end{equation}
Figure\,\ref{fig:config} shows the D7-brane configurations in the $y$-$\rho$ plane with $\psi=0$. Note that the $w$ direction is not shown since it is perpendicular to $y$ and $\rho$ directions in Fig.\,\ref{fig:config}. The Minkowski embedding is denoted by dotted line in Fig.\,\ref{fig:config}.
In the presence of the charge density and the magnetic field, there are two phases which are classified by the D7-brane configuration in black hole embedding. One is a trivial solution $\theta(u)=0$\,(denoted by solid line in Fig.\,\ref{fig:config}). From Eqs.\,(\ref{eq:theta}) and (\ref{eq:qq}), the chiral condensate of the dual field theory obviously becomes zero in this phase. The other is a non-trivial solution $\theta(u)\neq 0$\,(denoted by dashed line in Fig.\,\ref{fig:config}), which has a finite value of the chiral condensate in the dual field theory. In terms of the dual field theory, there are two phases which are classified by the chiral symmetry:\,the chiral symmetry restored\,($\chi$SR) phase and the chiral symmetry broken\,($\chi$SB) phase, respectively. To summarize, there are three types of solutions:\,the meson bound state with $\chi$SB, the meson melting state with $\chi$SB, and the meson melting state with $\chi$SR.

\begin{figure}[tbp]
	\centering
	\includegraphics[width=12cm]{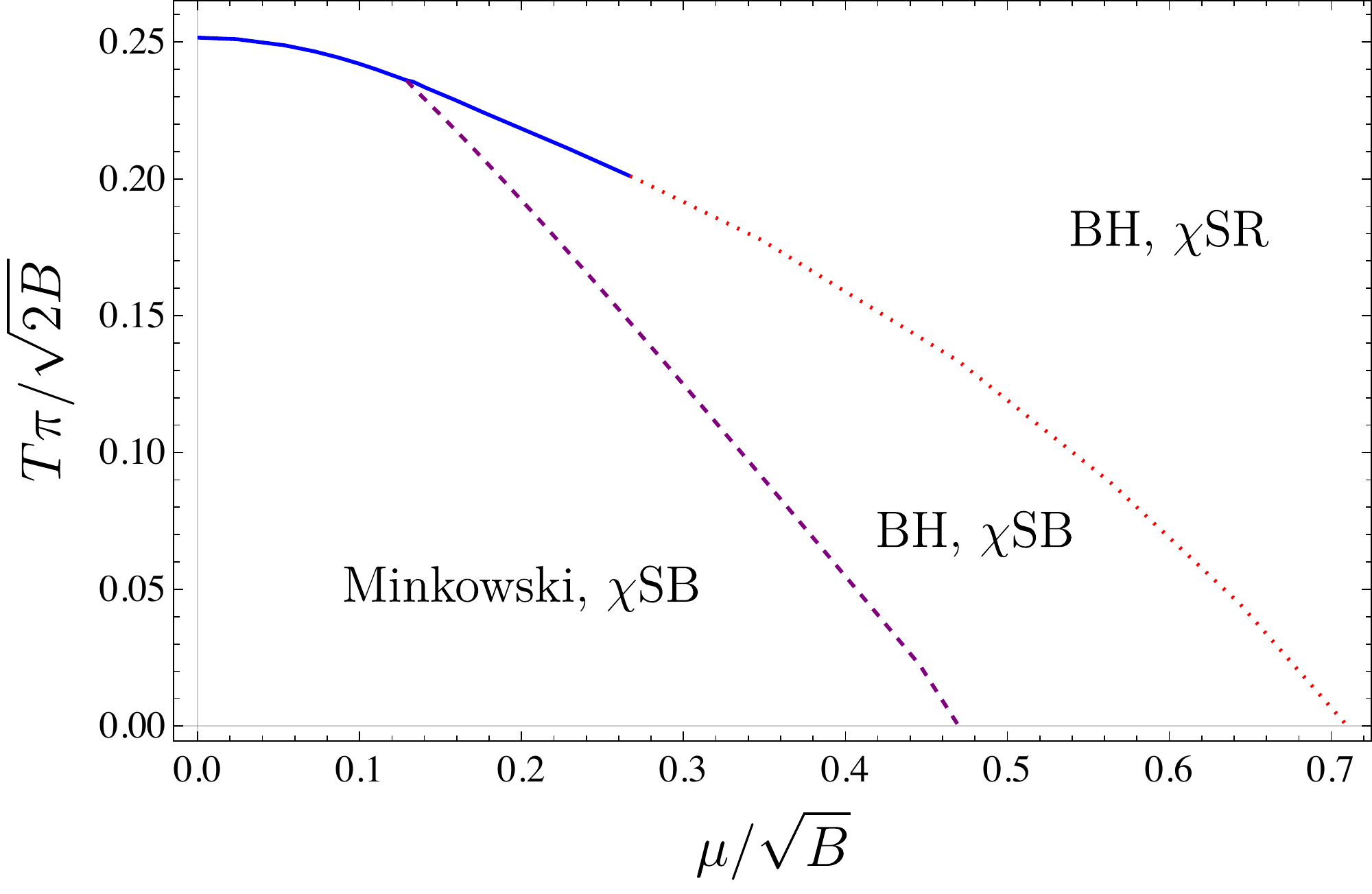}
	\caption{The phase diagram of our system. The parameters are temperature and chemical potential. There are three phases:\,Minkowski embedding, black hole embedding\,(denoted by BH in the phase diagram) with non-trivial configuration\,(which corresponds to chiral symmetry breaking phase, ${\rm \chi SB}$, in the dual field theory), and black hole embedding with trivial configuration\,(which corresponds to chiral symmetry restored phase, ${\rm \chi SR}$, in the dual field theory.). These phases are separated by 1st order phase transition\,(solid curve) and 2nd order phase transitions\,(dashed and dotted curves).}
	\label{fig:phase}
\end{figure}

These three types of solutions are realized by controlling parameters of $T$ and $D$. Each point of the phase transition between these phases are determined by the thermodynamic potential as studied in \cite{Evans2010}. The thermodynamic potential of the boundary theory is defined from the on-shell action by
\begin{equation}
	\Omega(T,\mu) \equiv \frac{-\tilde{S}_{D7}}{T_{D7} {\rm Vol}}=\int_{0}^{u_{H}} du \frac{\cos^{6} \theta}{u^{5}}(1+B^{2}u^{4})\sqrt{\frac{1+(1-u^{4}/u_{H}^{4} )u^{2} \theta'^{2}}{D^{2} u^{6} +(1+B^{2}u^{4})\cos^{6}\theta}},
\end{equation}
where ${\rm Vol}$ is the volume integral in 7-dimensional spacetime except for the radial space $u$. The chemical potential $\mu$ is defined by
\begin{equation}
	\mu \equiv \lim_{u \to 0} a_{t}(u)=\int_{0}^{u_{H}}du  \,D \sqrt{\frac{1+(1-u^{4}/u_{H}^{4} )u^{2} \theta'^{2}}{D^{2} u^{6} +(1+B^{2}u^{4})\cos^{6}\theta}}.
\end{equation}
The phase diagram in the parameter space of $\mu/ \sqrt{B}$ and $\pi T/\sqrt{2B}$ is shown in Fig.\,\ref{fig:phase}.
As we mentioned, there are three types of phases which are classified by the configurations of the D7-brane and the chiral symmetry of the dual massless theory. These phases are separated by 1st order phase transition\,(solid line) and 2nd order phase transition\,(dashed and dotted lines). At large chemical potential and/or high temperatures, the chiral symmetry is preserved. If we control these parameters, however, the chiral symmetry is spontaneously broken as shown in the middle and left regions of the phase diagram. The right and middle phases correspond to the meson melting phase with finite charge density, which implies that the system is dissipative. It is known that the meson melting phase with no charge density is unstable by computing the thermodynamic potential\,\cite{Evans2011mu,Albash:2007bk}.  In this paper, our interest is to study the NG mode which should appear when the chiral symmetry is spontaneously broken in the dual field theory.

\section{Nambu-Goldstone mode in equilibrium} \label{sec3}
In order to study the NG mode associated with the spontaneous chiral symmetry breaking, we focus only on black hole embedding. In the picture of the gravity theory, the broken symmetry corresponds to the $SO(2)$ rotational symmetry. The $SO(2)$ group rotates two directions $(y,w)$ which are perpendicular to the D7-brane. For instance, the $SO(2)$ rotational symmetry is broken in the middle bending configuration in Fig.\,\ref{fig:config}. As mentioned above, this solution corresponds to the chiral symmetry breaking phase in the dual field theory. On the other hand, the $SO(2)$ rotational symmetry is preserved in the lower flat configuration in Fig.\,\ref{fig:config}, which corresponds to the chiral symmetry restored phase. Since $\psi$ denotes the angular coordinate in the $y$-$w$ plane, the fluctuation with respect to the $\psi$ direction in the $SO(2)$ broken phase should correspond to the NG mode. The dual description of the NG mode is the fluctuation of the phase of the complex scalar field\,\cite{Myers2007}. After one takes the decoupling limit, its expectation value gives the mass of the hypermultiplet.

\subsection{Fluctuation }
The Wess-Zumino term provides a coupling between \(\psi\) and \(A_{a}\).
The relevant term is given by
\begin{equation}
	S_{\mathrm{WZ}} \simeq
	\mathrm{Vol}_{\mathrm{S}^3}\frac{T_{\mathrm{D7}}}{8}\int d^{4}x d u
	\epsilon^{ijklm} \cos^4\theta \partial_i \psi F_{jk} F_{lm},
\end{equation}
where \(i,j,k \cdots\) denote \(\sigma^{i}= t, x, y, z, u\).
If we consider the fluctuation with respect to the $\psi$ direction, it couples to fluctuations of the $U(1)$ gauge field via the Wess-Zumino term in the presence of the magnetic field.\footnote{Considering this coupling between the fluctuations, the dispersion relation at zero temperature was studied in \cite{Filev2009}. }
If we consider only the perpendicular direction of the fluctuation momentum with respect to the magnetic field direction\,($z$ direction), only the fluctuation of the gauge field $A_{z}$ couples to $\psi$. For simplicity, we focus only on these two fluctuations.
Since we set $\psi=0$ in the background setup, we consider the following fluctuations:
\begin{equation}
	\psi=0+\delta\psi(t,x,y,u), \hspace{1em}  A_{z}=\delta A_{z}(t,x,y,u).
\end{equation}
Expanding the action up to quadratic order in $\delta\psi$ and ${\delta A}_{z}$, we obtain the equations of motion for the fluctuations:
\begin{equation}
\begin{gathered}
	\partial_{a}\sqrt{-\det (g+F)} \gamma^{ab}g_{\psi\psi}\partial_{b}\delta\psi-(\cos^{4}\theta)' B \partial_{t} {\delta A}_{z} = 0, \\
	-2\partial_{a}\sqrt{-\det (g+F)} \gamma^{ab}g^{zz}\partial_{[b}{\delta A}_{z]}+(\cos^{4}\theta)' B \partial_{t} \delta\psi =0,
\end{gathered}
\label{eq: linearized eom}
\end{equation}
where $\gamma^{ab}$ is the inverse of the open string metric defined by $\gamma_{ab}=g_{ab}-(F g^{-1}F)_{ab}$.
(See appendix \ref{apdx: osm}).
Since \(\psi\) has singular behavior when \(y=0\), we shall use \(w\).
From (\ref{eq:coord}), the fluctuations are related by
\begin{equation}
	w \delta w + y \delta y = \frac{\sin \theta \cos \theta}{u^2} \delta \theta,~~
	y \delta w - w \delta y = \frac{\sin^2 \theta}{u^2} \delta \psi.
\end{equation}
In the current case, we can write \(\delta \psi = \delta w/y\) by setting \(w=0\) and \(y=\sin\theta/u\).%
\footnote{
	Actually, we cannot avoid the coordinate singularity by this replacement after writing the equation of motion.
	In order to avoid the singularity at \(m_q=0\), we must derive the equation of motion by employing a Cartesian coordinate.
	Alternatively, we consider small mass for the massless cases to avoid the singularity.
	We have checked our calculation gives same results compared to the result working in the Cartesian coordinates for small mass.
}
%For \(\delta w\) and \({\delta A}_{z}\), (\ref{eq: linearized eom}) are solvable even when \(y(u)\) reaches to zero.
In our setup, the non-trivial components of the open string metric are $\gamma^{tt}$ and $\gamma^{uu}$:
\begin{equation}
	\gamma^{tt}=-\frac{u^{2}}{f(u)} \frac{1+u^{2}f(u)\theta'(u)^{2}}{1+u^{2}f(u)\theta'(u)^{2}-u^{4} a_{t}'(u)^{2}}, \hspace{1em} \gamma^{uu}=\frac{u^{2}f(u)\theta'(u)^{2}}{1+u^{2}f(u)\theta'(u)^{2}-u^{4} a_{t}'(u)^{2}}.
\end{equation}
Here, we employ the plane wave ansatz for the fluctuations:
\begin{equation}
	\delta w(t,x,y,u) = {\cal W}(u) e^{-i\omega t +i \vec{k} \cdot \vec{x}} , \hspace{1em}
	\delta A_{z} = {\cal A}_{z}(u) e^{-i \omega t +i \vec{k}\cdot\vec{x}},
\end{equation}
where $\vec{k}=(k_{x},k_{y},k_{z})$ and $\vec{x}=(x,y,z)$ in our setup. As mentioned above, we assume $k_{z}=0$ in this paper for simplicity. In order to obtain the physical solutions, we impose the following boundary condition for fluctuations. Near the black hole horizon, we impose the ingoing-wave boundary condition:
\begin{equation}
	{{\cal W}}(u)=f(u)^{-i\frac{\omega}{4\pi T}}\tilde{\cal W}(u), \hspace{1em} {\cal A}_{z}(u)=f(u)^{-i\frac{\omega}{4\pi T}}\tilde{\cal A}_{z}(u).
\end{equation}
Recall that \(f(u)\) has a leading term of \(\order{u-u_H}\) in the vicinity of \(u=u_H\), this field-redefinition represents the Frobenius expansion at \(u=u_H\).
In the vicinity of \(u=0\), these fields have the following solutions:
\begin{equation}
	\tilde{\cal W} = {\cal W}^{(0)} + {\cal W}^{(2)} u^2 + \cdots,~~
	\tilde{\cal A}_{z} = {\cal A}_{z}^{(0)} + {\cal A}_{z}^{(2)} u^2 + \cdots.
\end{equation}
In the both cases, the first term is the non-normalizable mode and the second is the normalizable mode.
At the AdS boundary, we assume that ${\cal W}(u=0)=0$ and ${\cal A}_{z}(u=0)=0$ to pick up the normalizable solutions. Introducing the dimensionless quantity
\(
	{\textswab w}= \omega/4\pi T,
\)
we can rewrite the equations of motion for $\tilde{{\cal W}}(u)$ and $\tilde{\cal A}_{z}(u)$ as
% To be fixed
\begin{subequations}
\begin{gather}
\begin{aligned}
	\partial_u\left[ {\cal F}_{w}\gamma^{uu} \left(
			\partial_{u} \frac{\tilde{\cal W}}{y}(u)
			-i {\textswab w}\frac{f'}{f}\frac{\tilde{\cal W}}{y}(u)
	\right) \right]
	-i{\textswab w} \frac{f'}{f} {\cal F}_{w}\gamma^{uu}\left(
		\partial_{u} \frac{\tilde{\cal W}}{y}(u)
		-i{\textswab w}\frac{f'}{f}\frac{\tilde{\cal W}}{y}(u)
	\right)&\\
	-\left(\omega^{2}\gamma^{tt}+k_{\perp}^{2} \gamma^{\perp\perp}\right){\cal F}_{w}\frac{\tilde{\cal W}}{y}(u) - \omega(\cos^{4}\theta)'B \tilde{\cal A}_{z}(u) &= 0,
\end{aligned}\\
\begin{aligned}
	\partial_u\left[ {\cal F}_{\cal A}\gamma^{uu} \left(
		\partial_{u} \tilde{\cal A}_{z}(u)
		-i {\textswab w}\frac{f'}{f}\tilde{\cal A}_{z}(u)
	\right) \right]
	-i{\textswab w} \frac{f'}{f} {\cal F}_{\cal A}\gamma^{uu}\left(
		\partial_{u}\tilde{\cal A}_{z}(u)-i{\textswab w}\frac{f'}{f}\tilde{\cal A}_{z}(u)
	\right)&\\
	-\left(\omega^2\gamma^{tt}+k_{\perp}^{2} \gamma^{\perp\perp}\right){\cal F}_{\cal A}\tilde{\cal A}_{z}(u) - \omega(\cos^{4}\theta)'B \frac{\tilde{\cal W}}{y}(u)&=0,
\end{aligned}
\end{gather}
\end{subequations}
where ${\cal F}_{w}=\sqrt{-\det(g+F)}g_{\psi\psi}$ and ${\cal F}_{\cal A} = \sqrt{-\det(g+F)} g^{zz}$. $\perp$ denotes the transverse directions\,($x$,\,$y$) with respect to the magnetic field. Here, since the background is isotropic with respect to $x$ and $y$ directions, $k_{\perp}$ denotes $k_{x}$ or $k_{y}$.
 In this paper, we employ the shooting method to numerically solve the equations of motion for fluctuations. Then, we calculate the eigenvalues $\omega$ as functions of $k_{\perp}$ so that the solutions satisfy the above two boundary conditions.

\subsection{Dispersion relation}
We analyze the behaviors of several QNMs. In our study, we focus on the lowest two modes.
%The frequency of the QNM vanishes at zero momentum in a gapless mode, while it becomes finite in a gapped mode.
We investigate the dispersion relation associated with these two modes in our anisotropic background.
We show the behavior of these two modes as a function of $k_{\perp}/\sqrt{B}$ for several values of $D/B^{3/2}$ with $\pi T/\sqrt{2B}$ fixed in Fig.\,\ref{fig:disp1}.%
\footnote{
	We also show more examples of the dispersion relation in appendix \ref{apdx:disp}.
}
\begin{figure}[htbp]
\centering
\includegraphics[width=7cm]{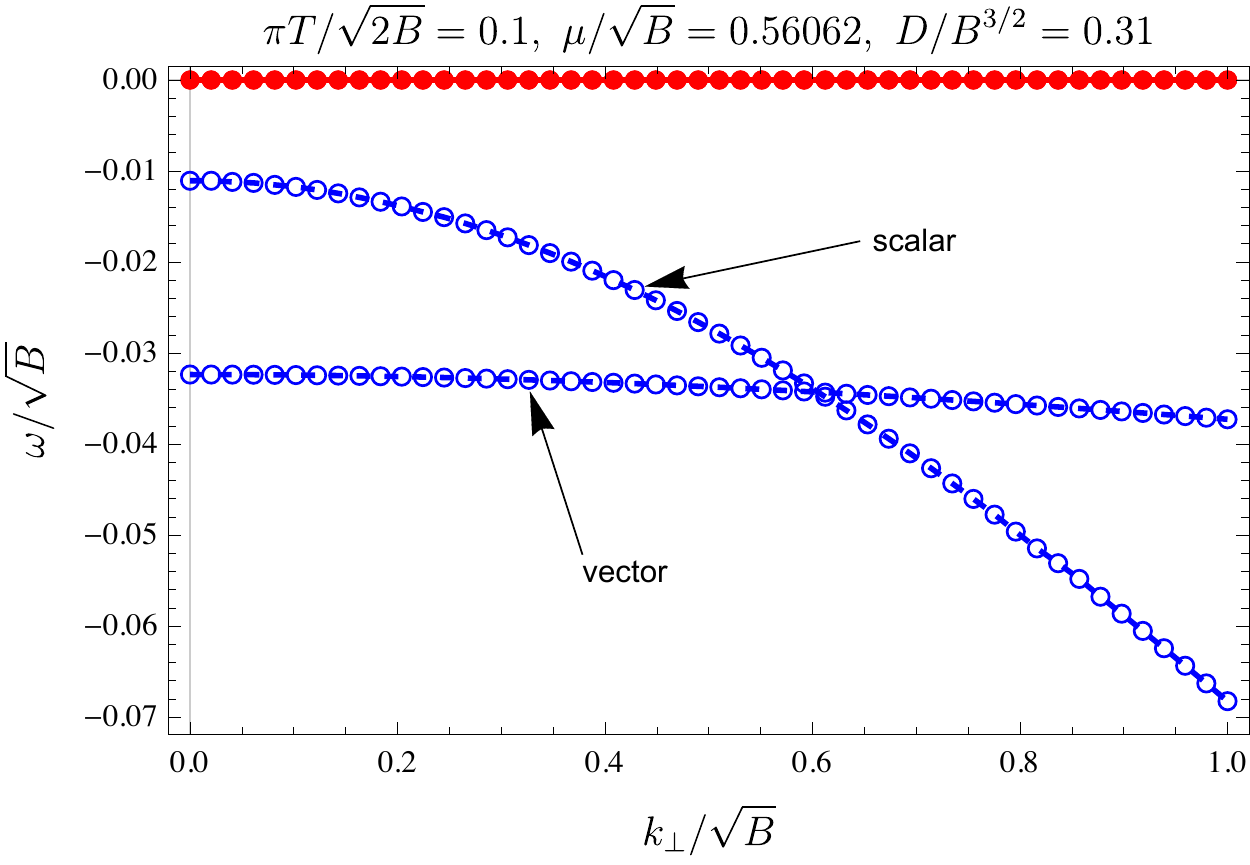}
\includegraphics[width=7cm]{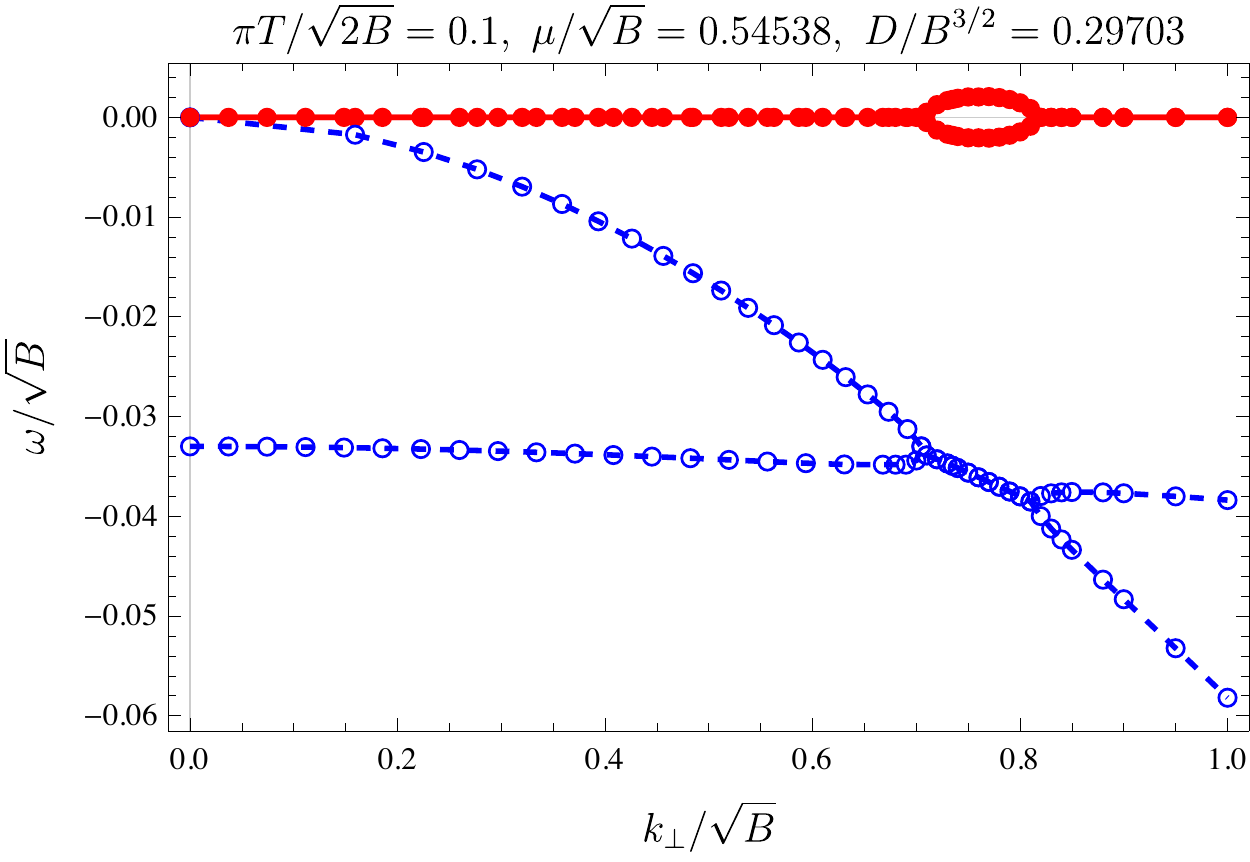}
\includegraphics[width=7cm]{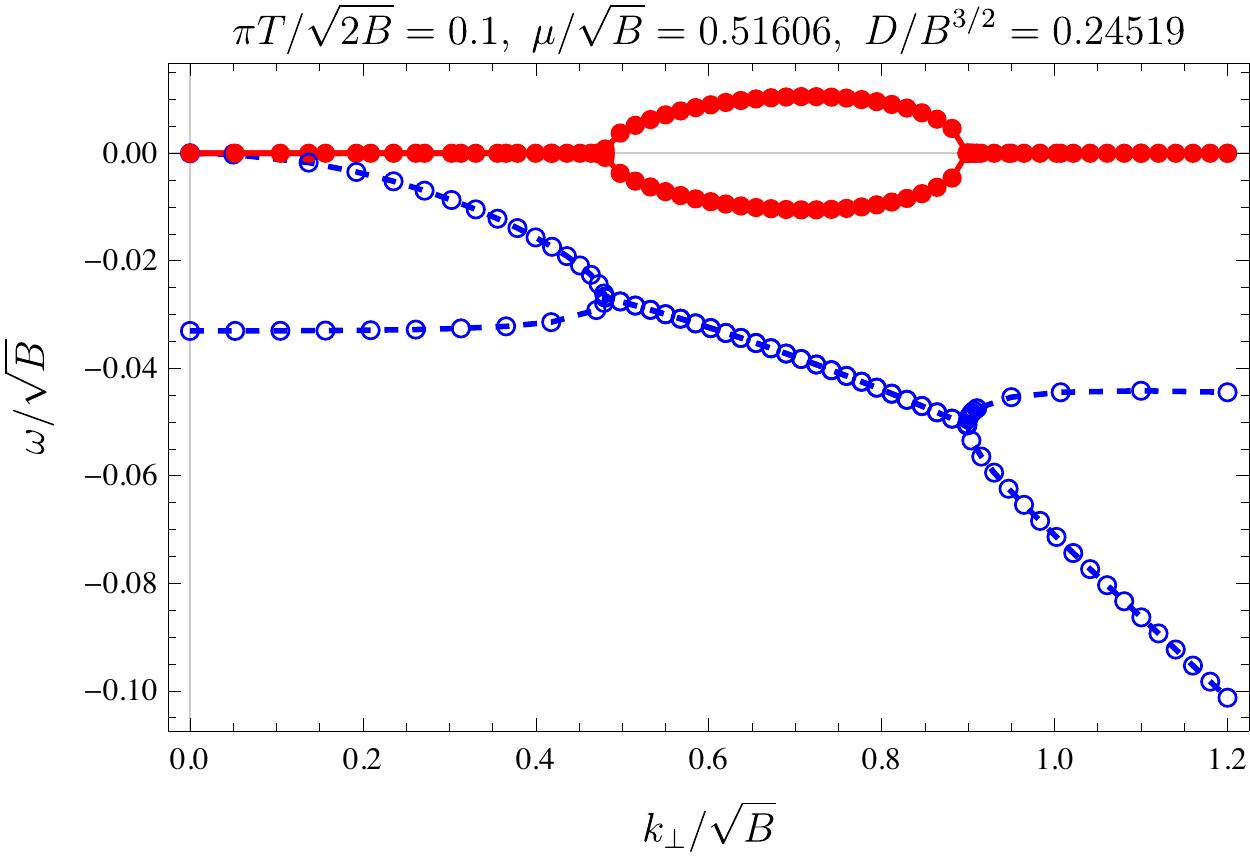}
\includegraphics[width=7cm]{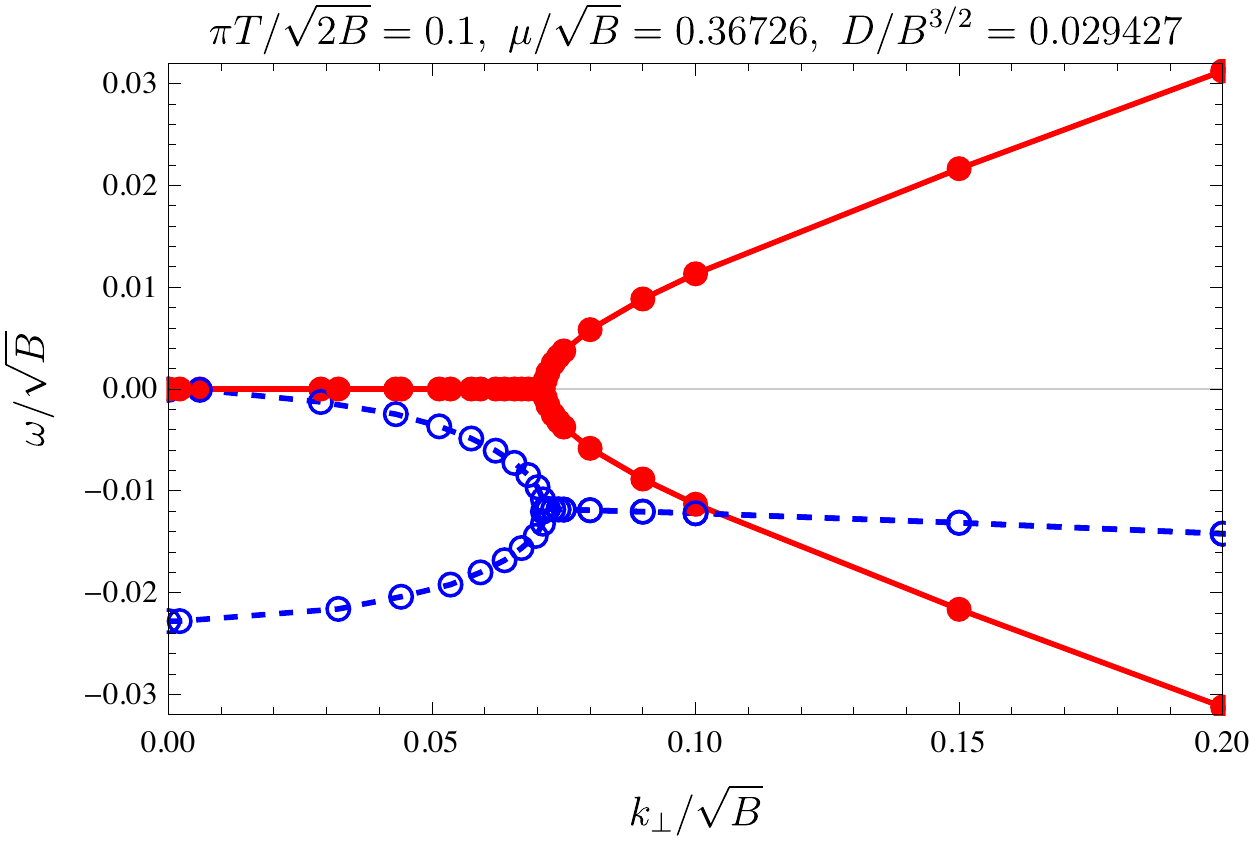}
\caption{The behavior of the real part\,(filled circles) and the imaginary part\,(open circles) of $\omega/\sqrt{B}$ as a function of $k_{\perp}/\sqrt{B}$.
In the upper left panel, the system preserves the chiral symmetry and there is no gapless mode. The fluctuations for the scalar field and the vector field\,(gauge field) are decoupled from each other.
In the other panels, the system spontaneously breaks the symmetry and the gapless mode arises.
}
\label{fig:disp1}
\end{figure}
In each plot, filled circles and open circles denote the real part and the imaginary part of $\omega/ \sqrt{B}$, respectively. In the chiral symmetry restored phase\,(upper left), the fluctuations of $\psi$\,(scalar) and $A_{z}$\,(vector) are decoupled from each other. We find that each fluctuation has a purely imaginary mode. In the chiral symmetry broken phase, however, these fluctuations are coupled with each other. For larger $D/B^{3/2}$\,(upper right or lower left), one can see that there are a gapless mode and a gapped mode. Note that the frequency of the QNM vanishes when the momentum becomes zero for a gapless mode, whereas it becomes finite for a gapped mode. As can be seen from Fig.\,\ref{fig:disp1}, the gapless mode behaves as the diffusive mode near zero momentum: $\omega=-iD_{\perp}k^{2}$ ($D_{\perp}$ is a real positive constant). This diffusive behavior of the NG mode in the dissipative system is consistent with that studied in Ref.\,\cite{Minami2018}. At specific momentum, the real parts of these two modes become non-zero at this momentum. In other words, the hydrodynamic diffusive behavior is changed into the reactive regime\footnote{More precisely, the fluctuations are changed from a purely damped decay regime to a slowly damped oscillation regime at that specific momentum. This behavior has been observed in the same model as discussed in\,\cite{Kaminski2009,Kaminski2009d}.} with the non-vanishing real part. As we increase $k_{\perp}/\sqrt{B}$, the one mode splits into two modes and the real parts of those modes become zero again. For smaller $D/B^{3/2}$\,(lower right), the dispersion relation for small momenta agrees with that derived from the {\it telegraph equation}:
\begin{equation}
	-\tau \omega^{2}- i \omega + D_{\perp} k_{\perp}^{2}=0,
	\label{eq:telegraph}
\end{equation}
where $\tau$ is the relaxation time of the corresponding mode. In fact, we find that the numerical plot is well-fitted by (\ref{eq:telegraph}) for \(0<k/\sqrt{B}<0.1\) with $\tau=43.114 \times 1/\sqrt{B}$ and $D_{\perp} = 1.1416 \times 1/\sqrt{B}$ as shown in Fig.\,\ref{fig:dispfit}.

\begin{figure}[htbp]
\centering
\includegraphics[width=12cm]{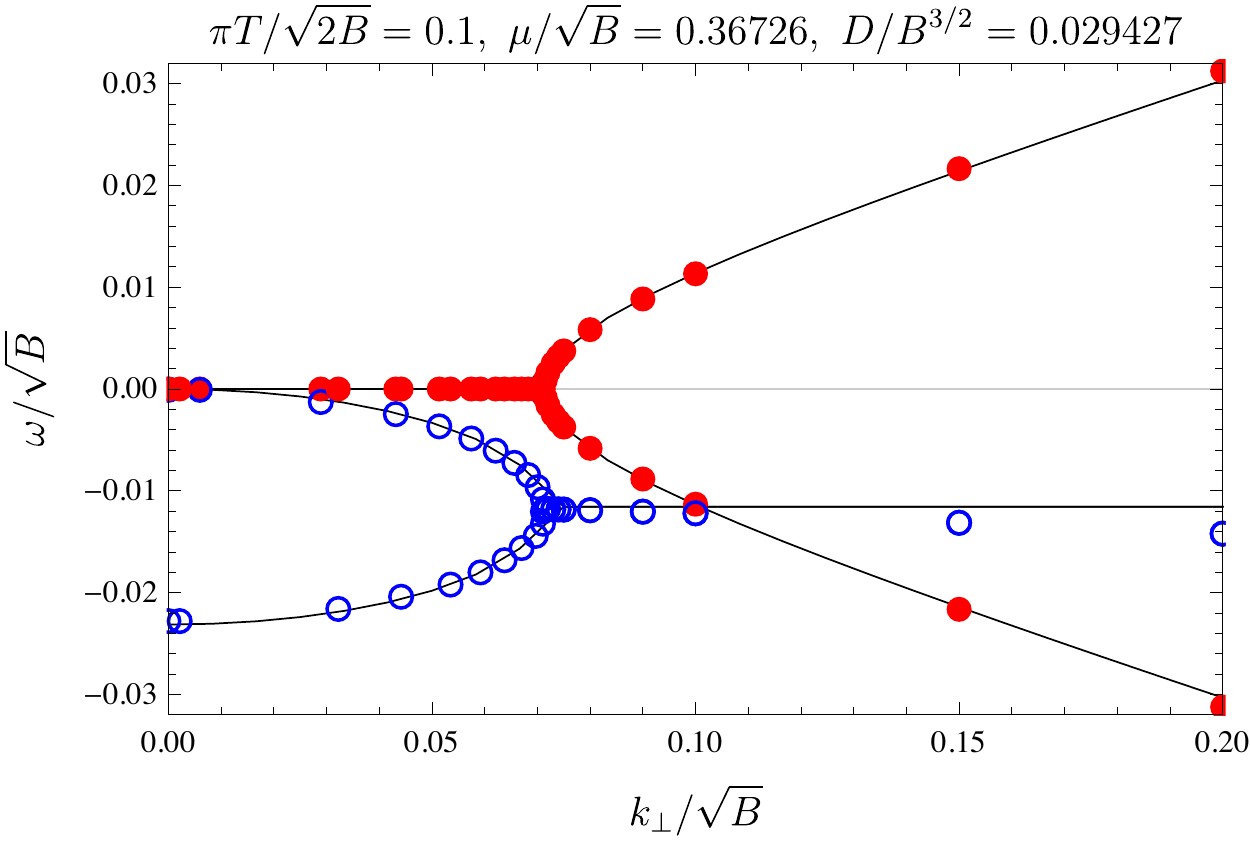}
\caption{The model fitting by (\ref{eq:telegraph}) for the dispersion relation in the lower-right panel of Fig.\,\ref{fig:disp1}.
The black solid lines are corresponding to the fitting curves with $\tau=43.114 \times 1/\sqrt{B}$ and $D_{\perp} = 1.1416 \times 1/\sqrt{B}$.
}
\label{fig:dispfit}
\end{figure}

Here, there are two different points:\,the merger point of two purely imaginary modes and the re-emerging point of them. The former can be generally observed in the dispersion relation of a dissipative system. In terms of the real part of the frequency, the dispersion relation has a gap in momentum space. For this reason, it is also referred to as the {\it k-gap}. The dispersion relation with the k-gap is widely observed in various physical systems\footnote{In the effective field theory approach, this type of NG mode was shown in a quantum time-crystal as a NESS\,\cite{Hayata2018}.}\,(for example, see Ref.\,\cite{Baggioli2019}). 
In terms of holography, the dispersion relation which agrees with the telegraph equation has been generally discussed as the diffusion-to-sound crossover in \cite{Grozdanov2018}.
In fact, the k-gap has been numerically obtained in various models\,\cite{Andrade2018,Baggioli2018v,Baggioli2018n,Jimenez-Alba2014,Arias2014,Gran2018,Song2018,Itsios2018}

On the other hand, the similar behavior to the latter has been observed in the holographic system in which the translation symmetry is broken\,\cite{Alberte2017,Baggioli2019b,Amoretti2019b}.
We consider the origin of this short wavelength behavior in our system is different from their systems because our system is spatially homogeneous, but there could be same mechanism, in behind.
We expect that the short wavelength behavior results from the scattering between charged particles in the presence of the charge density since the re-emerging point of the purely imaginary modes will become to \(k_{\perp} \rightarrow \infty\) in the limit of \(D \rightarrow 0\).%
\footnote{
	In \cite{Yang2017}, the similar behavior at large momentum were also observed in the molecular dynamics simulations. This beahvior is related to the effect of inter-atomic separation in this stduy.
}

\begin{figure}[htbp]
\centering
\includegraphics[width=15cm]{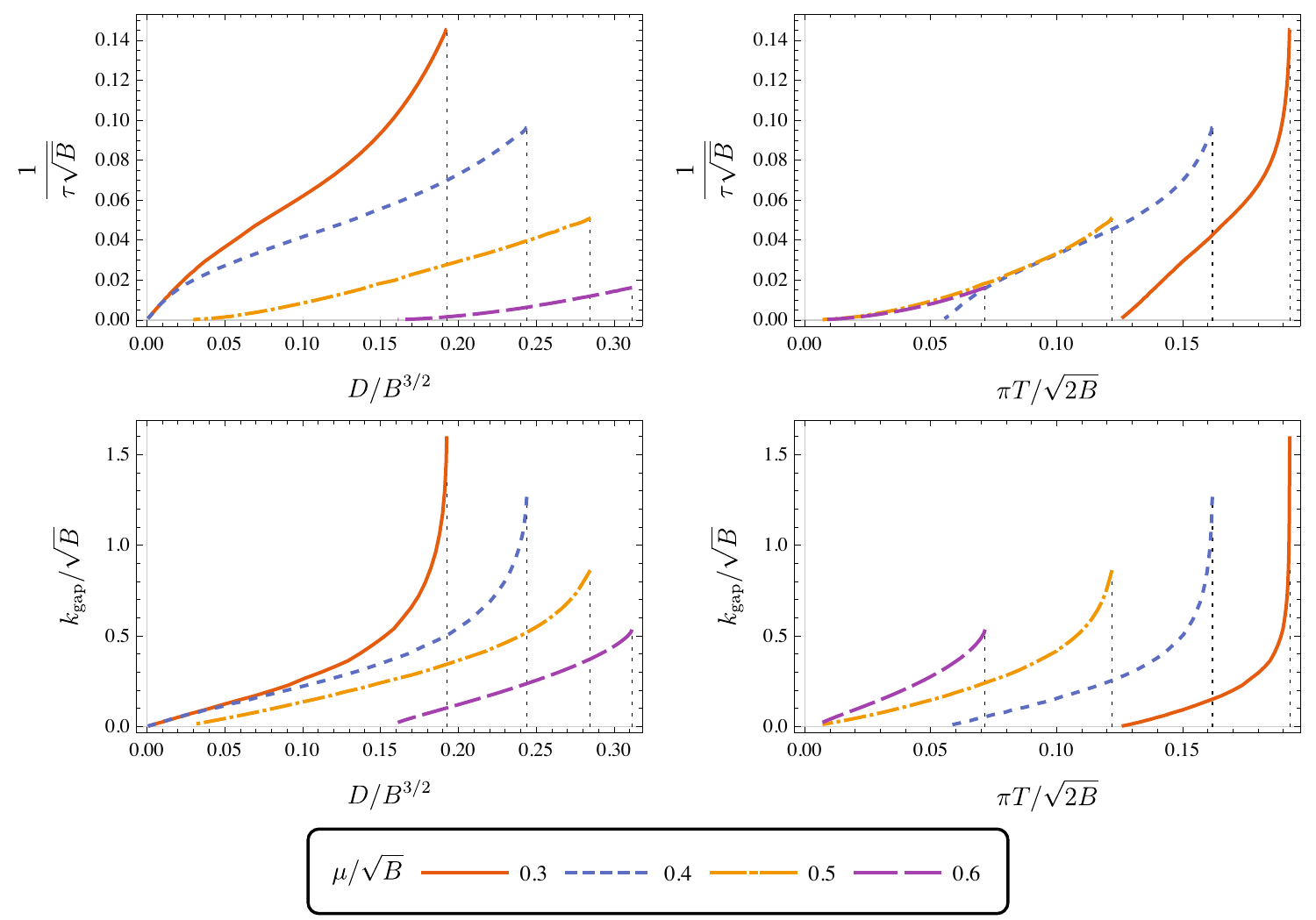}
\caption{The behaviors of $1/(\tau\sqrt{B})$ and $k_{\rm gap}/\sqrt{B}$ as a function of $D/B^{3/2}$\,(left panel) and the behaviors of $1/(\tau\sqrt{B})$ and $k_{\rm gap}/T$ as a function of $\pi T/ \sqrt{2B}$\,(right panel) for several values of $\mu/\sqrt{B}$.}
\label{fig:taukgap}
\end{figure}

In the last of this section, we further study the properties of the relaxation time $\tau$ and the k-gap momentum $k_{\rm gap}$ with respect to each parameter. Here, we define $k_{\rm gap}$ as the specific smallest momentum at which the real parts of a gapless mode and a gapped mode become non-zero. We show their behaviors as a function of $D/B^{3/2}$ and $\pi T / \sqrt{2B}$ for several values of $\mu/\sqrt{B}$ in Fig.\,\ref{fig:taukgap}.
Note that there are minimum values of $D/B^{3/2}$ and $\pi T / \sqrt{2B}$ in each plot since the system undergoes a transtion to Minkowski embedding below these values. We find that both the inverse of the relaxation time and the k-gap momentum monotonically increase with increasing $D/B^{3/2}$ and $\pi T / \sqrt{2B}$.
In particular, $k_{\rm gap}/\sqrt{B}$ is linearly increasing for small $D/B^{3/2}$ in each plot.
It also turns out that $1/(\tau \sqrt{B})$ is increasing as square power of \(\pi T/\sqrt{2B}\) at low temperatures.
This means that \(1/\tau T\) linearly depends on \(\pi T/\sqrt{2B}\) at low temperatures.%
\footnote{
	Due to the numerical instabilities, we could not get data for sufficiently small \(\pi T/\sqrt{2B}\).
	%The data at low temperatures, however, tend to exhibit the linear dependence on \(\pi T/\sqrt{2B}\).
}
Our result agrees with the observations in \cite{Baggioli2018v,Baggioli2018n}, which show the relaxation time has simple $1/T$ dependence in liquids and the corresponding holographic model. In our case, however, it is no longer linear for smaller $\mu/\sqrt{B}$ and/or larger $\pi T/\sqrt{2B}$. 
%We also study the relationship between the relaxation time and the k-gap momentum in Fig.\,\ref{fig:tauVSkgap}.
%\begin{figure}[htbp]
%\centering
%\includegraphics[width=13cm]{kgap_vs_tau}
%\caption{The plot of $k_{\rm gap}/\sqrt{B}$ as a function of $1/\tau\sqrt{B}$ for several values of $\mu/\sqrt{B}$}
%\label{fig:tauVSkgap}
%\end{figure}
%One can see that the k-gap momentum increases if the relaxation time decreases. This can be understood from the fact that both the k-gap momentum and the inverse of the relaxation time increase with increasing temperature. At small value of $\mu/\sqrt{B}$, the dependence can be linear for sufficiently smaller $1/\tau\sqrt{B}$, which implies that the dispersion relation agrees with that in the telegraph equation.

\section{Nambu-Goldstone mode in non-equilibrium steady state} \label{sec4}
In this section, we analyze the NG mode in the presence of the constant current. Since the constant current is induced by the external electric field, the system is in the NESS, namely $J \cdot E\neq0$. We show that the dispersion relation of the NG mode in the NESS system exhibits a linear dispersion in the real part of the QNM frequency in addition to the diffusive behavior. Moreover, we analyze the linear dispersion of the NG mode in the hydrodynamic approximation.

\subsection{Setup}
We shortly review our setup in which there is a constant current in the presence of the external electric field in addition to the chemical potential and the magnetic field. Here, we consider that the direction of the electric field is perpendicular to the magnetic field. Therefore, we assume that the gauge fields are given by\,\cite{Ammon2009,Evans2011mu}
\begin{equation}
	A_{t}(u)=a_{t}(u), \hspace{1em} A_{x}(t,u)=-Et+a_{x}(u), \hspace{1em} A_{y}(x,u)=Bx+a_{y}(u).
\end{equation}
Since the D7-brane action depends only on the $u$ derivatives of $A_{t}$, $A_{x}$, and $A_{y}$, there are three conserved quantities:
\begin{equation}
	D=\frac{\delta S}{\delta A_{t}'}, \hspace{1em} J_{x}=\frac{\delta S}{\delta A_{x}'}, \hspace{1em} J_{y}=\frac{\delta S}{\delta A_{y}'}.
\end{equation}
Here, $J_{x}$ and $J_{y}$ are identified as the components of $U(1)_B$ current density in the dual field theory\,\cite{OBannon2007}. Given these conserved quantities, we can obtain the gauge fields solutions:
\begin{subequations}
\begin{eqnarray}
	a_{t}'(u)&=&-\frac{\sqrt{-g_{tt}g_{uu}}}{g_{xx}^{2}}\frac{\xi J_{t} - B a}{\sqrt{\xi \chi -\frac{a^{2}}{g_{xx}^{2}}}}, \\
	a_{x}'(u)&=& \sqrt{\frac{-g_{uu}}{g_{tt}}}\frac{1}{g_{xx}} \frac{\xi J_{x}}{\sqrt{\xi \chi -\frac{a^{2}}{g_{xx}^{2}}}}, \\
	a_{y}'(u)&=& \sqrt{\frac{-g_{uu}}{g_{tt}}}\frac{1}{g_{xx}} \frac{\xi J_{y}+E a}{\sqrt{\xi \chi -\frac{a^{2}}{g_{xx}^{2}}}},
\end{eqnarray}
\label{eq:gauge_sol}
\end{subequations}
where
\begin{subequations}
\begin{eqnarray}
	\xi &=& -g_{tt}g_{xx}^{3}-g_{tt}g_{xx}B^{2}-g_{xx}^{2}E^{2}, \label{eq:xi}\\
	\chi &=& -g_{tt}g_{xx}^{2} \cos^{6}\theta -J_{x}^{2}-J_{y}^{2} -\frac{g_{tt}}{g_{xx}}D^{2}, \\
	a &=& g_{xx}^{2}E J_{y}-g_{tt}g_{xx}B D.
\end{eqnarray}
\end{subequations}
For convenience, we perform the Legendre transformation, and we obtain the Legendre-transformed on-shell action
\begin{eqnarray}
	\hat{S}_{D7} &=& S_{D7} - \int d^{8}\sigma\left( F_{ut} \frac{\delta S}{\delta F_{ut}}+ F_{ux} \frac{\delta S}{\delta F_{ux}} +F_{uy} \frac{\delta S}{\delta F_{uy}} \right)  \\
	&=& -T_{D7} \int d^{8} \sigma \sqrt{-\frac{g_{uu}}{g_{tt}}}\frac{1}{g_{xx}} \sqrt{\xi \chi-\frac{a^{2}}{g_{xx}^{2}}}.
\end{eqnarray}
Since $\xi$ is negative at the horizon but positive at the boundary, $\xi$ must vanish at some value of $u$. We denote this as $u_{*}$ and it is referred to as the {\it effective horizon}. The condition of $\xi(u_{*})=0$ gives the explicit form of the $u_{*}$ as a function of $u_{H}$, $E$, and $B$.
%In the presence of an electric field $E$, the effective horizon emerges and its location $u_{*}$ is determined by the open string metric as discussed in Ref.\,\cite{Kim2011}.
%\begin{equation}
%	\det\gamma_{\alpha\beta}(u_{*})=0 \hspace{1em} \Longleftrightarrow \hspace{1em} g_{tt}g_{xx}^{2}+g_{tt}B^{2}+g_{xx}E^{2}=0,
%\end{equation}
%where $\alpha,\beta$ denote the spacetime coordinates of the boundary theory. The explicit components of the open string metric are shown in Appendix\,\ref{apdx: osm}.
The location of the effective horizon is explicitly written as
\begin{equation}
	u_{*}^{4}=\frac{1}{b^{2}}\left(\frac{b^{2}-e^{2}-1}{2}+\sqrt{\left(\frac{b^{2}-e^{2}-1}{2}\right)^{2}+b^{2}} \right)u_{H}^{4},
	\label{eq:zstar}
\end{equation}
where $e$ and $b$ are the dimensionless quantities defined by
\begin{equation}
	e=\frac{E}{\pi^{2}T^{2}}, \hspace{2em} b=\frac{B}{\pi^{2}T^{2}}.
\end{equation}

In order for the on-shell action to remain real for $0<u<u_{H}$, three functions $\xi$, $\chi$, and $a$ must vanish at $u_{*}$ as discussed in Ref.\,\cite{Ammon2009}. This condition gives us the currents $J_{x}$ and $J_{y}$. To be precise, the equation of $a(u_{*})=0$ gives us $J_{y}$ and we find $J_{x}$ from $\chi(u_{*})=0$ by substituting $J_{y}$. Thus, $J_{x}$ and $J_{y}$ are given by
\begin{eqnarray}
	J_{x}&=& \frac{g_{xx} E}{g_{xx}^{2}+B^{2}} \sqrt{ {\cal N}^{2} g_{xx}(g_{xx}^{2}+B^{2}) \cos^{6}\theta +D^{2}}, \\
	J_{y} &=& -\frac{BDE}{g_{xx}^{2}+B^{2}},
\end{eqnarray}
where all functions of $u$ are evaluated at the effective horizon $u=u_{*}$.

Since the effective horizon is located at the outside of the black hole horizon, we impose the boundary condition at $u=u_{*}$, instead of $u=u_{H}$. The boundary condition of $\theta(u_{*})$ is explicitly given by $\theta(u_{*})=\theta_{0}$. Here, we set the value of $\theta_{0}$ by hand in numerical calculations. Demanding the regularity for the solution, the boundary condition of $\theta'(u_{*})$ is obtained from the equation of motion for $\theta(u)$ at $u=u_{*}$. As explained in the previous section, we numerically solve the equation of motion for $\theta(u)$ by shooting method from the effective horizon to the AdS boundary. In the NESS system, there are two solutions of black hole embedding as with the equilibrium system:\,a trivial solution of $\theta(u)=0$ and a non-trivial solution of $\theta(u) \neq 0$. We note again that these solutions correspond to the chiral symmetry restored phase and the breaking phase in the dual field theory, respectively. In\,\cite{Evans2011mu}, it has been shown that the chiral symmetry is spontaneously broken in the presence of the external electric field.\footnote{Also in\,\cite{Imaizumi2019}, the phase diagram in the $J$-$B$ plane associated with the chiral symmetry breaking was studied in the NESS system without the charge density.} In the following section, we will study the NG mode in this NESS background.

\subsection{Dispersion relation}
As we analyzed in the previous section, we investigate the behavior of a gapless mode and a gapped mode. In the presence of an external electric field along $x$ direction, the non-equilibrium background is completely anisotropic.
Finite \(E\) gives additional terms to the linearized equations via the WZ term and the off-diagonal components of \(\gamma^{ab}\).
The equations of motion for the fluctuations are given by
\begin{equation}
\begin{alignedat}{3}
	\partial_a& \sqrt{-\det(g+F)}\gamma^{ab}g_{\psi\psi}\partial_b \delta \psi &
	+& (\cos^4\theta)'
	\left(
		E \partial_{y} \delta {A}_{z} - B \partial_{t} \delta {A}_{z}
	\right) &
	&= 0,\\
	- 2 \partial_a& \sqrt{-\det(g+F)} \gamma^{ab} g^{zz} \partial_{[b} \delta {A}_{z]}  &
	-& (\cos^4 \theta)' \left(
		E \partial_y \delta \psi - B \partial_t \delta \psi
	\right) &
	&=0.
\end{alignedat}
\label{eq: linearized eom E}
\end{equation}
As in the previous analysis, we use \(\delta w = y \delta \psi\).
Let us assume the plane wave ansatz:
\(
	\delta w = {\cal W}(u) e^{-i\omega t + i\vec{k}\cdot \vec{x}},~
	\delta A_{z} = {\cal A}_{z}(u) e^{-i\omega t + i\vec{k}\cdot \vec{x}},
\) where \(\vec{k}=(k_x, k_y, 0)\) and \(\vec{x}=(x,y,z)\).
The equations of motion are written as
%To be fixed
\begin{subequations}
\begin{gather}
\begin{aligned}
	\partial_{u} \Bigg[
		{\cal F}_{w}
		\left(
			\gamma^{uu} \partial_u \frac{\mathcal{W}}{y} (u)
			-ik_{\alpha}\gamma^{u\alpha} \frac{\mathcal{W}}{y}(u)
		\right)
	\Bigg]
	&- k_{\alpha} k_{\beta} {\cal F}_{w}
	\gamma^{\alpha\beta} \frac{\mathcal{W}}{y}(u)\\
	&+i\left(
		k_{y} E + \omega B
	\right)
	(\cos^4\theta)' \mathcal{A}_z(u)=0,
\end{aligned}\\
\begin{aligned}
	\partial_{u} \Big[
		{\cal F}_{\cal A}
		\left(
			\gamma^{uu} \partial_{u} \mathcal{A}(u)
			-ik_{\alpha}\gamma^{u\alpha} \mathcal{A}(u)
		\right)
	\Big]
	&- k_{\alpha}k_{\beta} {\cal F}_{\cal A}
	\gamma^{\alpha\beta} \mathcal{A} (u)\\
	&+i\left(
		k_{y} E + \omega B
	\right) (\cos^4\theta)' \frac{\mathcal{W}}{y}(u)=0,
\end{aligned}
\end{gather}
\end{subequations}
where ${\cal F}_{w}=\sqrt{-\det(g+F)}g_{\psi\psi}$ and ${\cal F}_{\cal A} = \sqrt{-\det(g+F)} g^{zz}$.
\(\alpha,\beta = t,x,y,z\) are indices of the boundary coordinates.
Here, we impose the ingoing-wave boundary conditions at the effective horizon and the vanishing Dirichlet condition at the AdS boundary. Near the effective horizon, the fluctuation fields can be written as
\begin{equation}
	{\cal W}(u)=(1-u/u_{*})^{\lambda}\hat{\cal W} (u), \hspace{1em} {\cal A}_{z}(u) =(1-u/u_{*})^{\lambda}\hat{\cal A}_{z}(u),
\end{equation}
where $\hat{\cal W}(u)$ and $\hat{{\cal A}_{z}}(u)$ are the regular parts at the effective horizon. At the effective horizon, we impose $\lambda=0$ which corresponds to the ingoing-wave condition for the fluctuations.\footnote{The explicit discussion for the ingoing-wave condition at the effective horizon is given in Refs.\,\cite{Mas2009, Ishigaki2020}}
At the AdS boundary, we assume that ${\cal W}(u=0)=0$ and ${\cal A}_{z}(u=0)=0$ as explained in the previous section.

\begin{figure}[tbp]
	\centering
	\includegraphics[width=14cm]{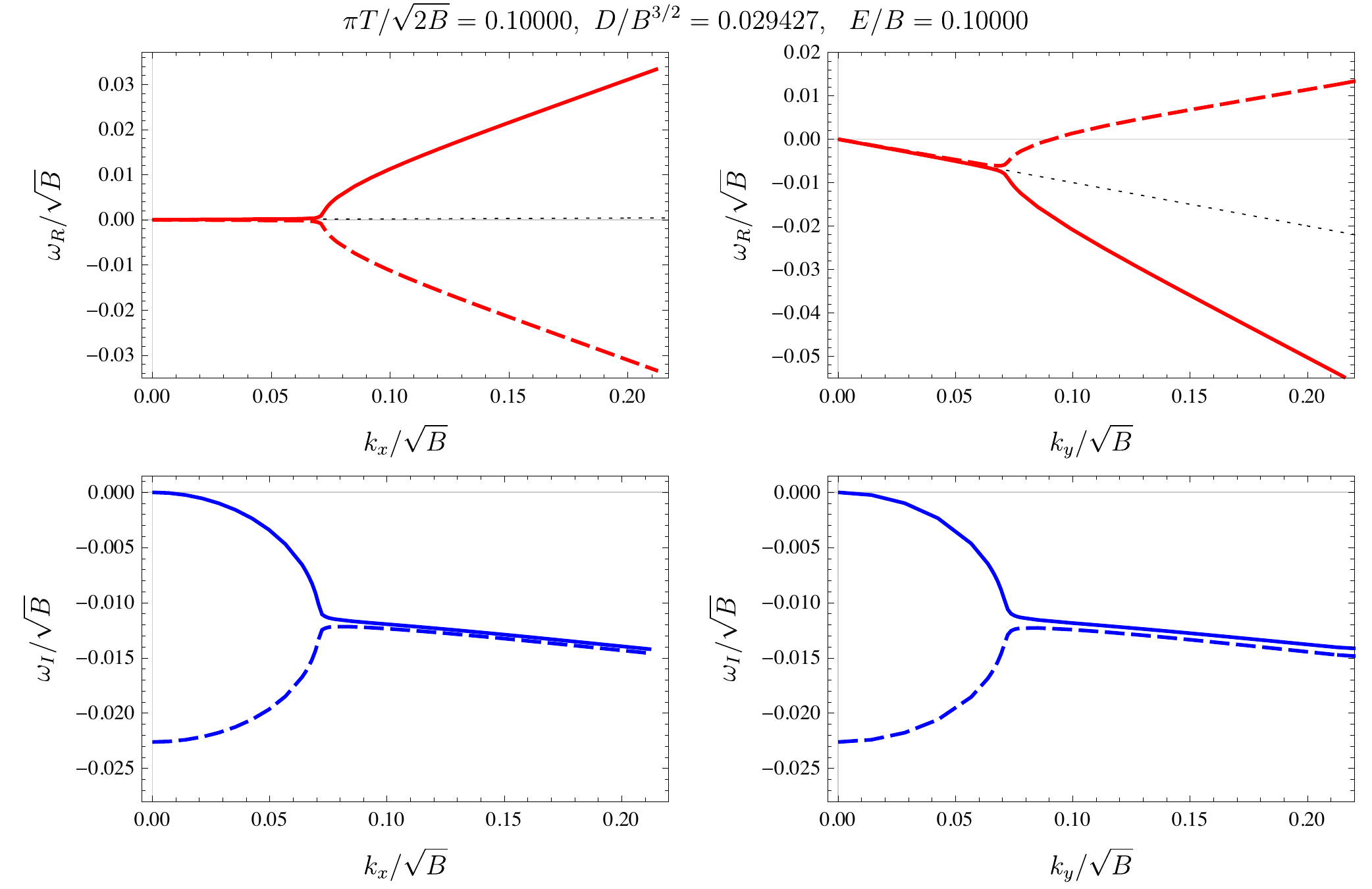}
	\caption{The behavior of the real part\,(upper panel) and the imaginary part\,(lower panel) of $\omega/\sqrt{B}$ as a function of $k_{x}/\sqrt{B}$\,(left panel) and $k_{y}/\sqrt{B}$\,(right panel) in the non-equilibrium state. The solid lines and the dashed lines denote the gapless modes and the gapped modes, respectively.
	In this configuration, $\beta_{x} = 0.0020007$ and $\beta_{y} = -0.099960$ in (\ref{eq:coefs}).
	The black dotted lines show \(\omega = \beta_{x} k_{x}\) and \(\omega = \beta_{y} k_y\) in the hydrodyanmic approximation, respectively.
	}
	\label{fig:dispNESS}
\end{figure}

We show a typical behavior of the real part\,(upper panel) and the imaginary part\,(lower panel) of $\omega/\sqrt{B}$ for a gapless mode\,(solid line) and a gapped mode\,(dashed line) as a function of $k_{x}/\sqrt{B}$ in the left panels of Fig.\,\ref{fig:dispNESS}. Here, note that we set $k_{y}=0$.
As in the equilibrium, we find that the gapless mode behaves as the diffusive mode near zero momentum. At a glance, the dispersion relation of these modes looks similar to that in equilibrium background as shown in Fig.\,\ref{fig:disp1}.
However, the real parts of these modes are sufficiently small but non-zero even in small momenta\,($k_{x}/ \sqrt{B} \lesssim 0.07$ in Fig.\,\ref{fig:dispNESS}). It appears that this difference arises from the anisotropic background as discussed later.
For finite \(k_y\), the behavior of the dispersion relation in the NESS background is dramatically changed.
We also show the behavior as a function of $k_{y}/\sqrt{B}$ with $k_{x}=0$ in the right panels of Fig.\,\ref{fig:dispNESS}.
One can see that these modes are no longer purely imaginary modes and the real parts of them become large even for small $k_{y}/\sqrt{B}$.
In other words, these modes represents the linear dispersion with respect to $k_{y}$:\,\(\omega = \beta_{y} k_y + \order{k_y^2}\) and \(\omega = -i/\tau + \tilde{\beta}_{y} k_y + \order{k_y^2}\) with a real-valued constant \(\beta_{y}, \tilde{\beta_{y}}\), and \(\tau\), in the hydrodynamic regime. 

\begin{figure}[tbp]
	\centering
	\includegraphics[width=14cm]{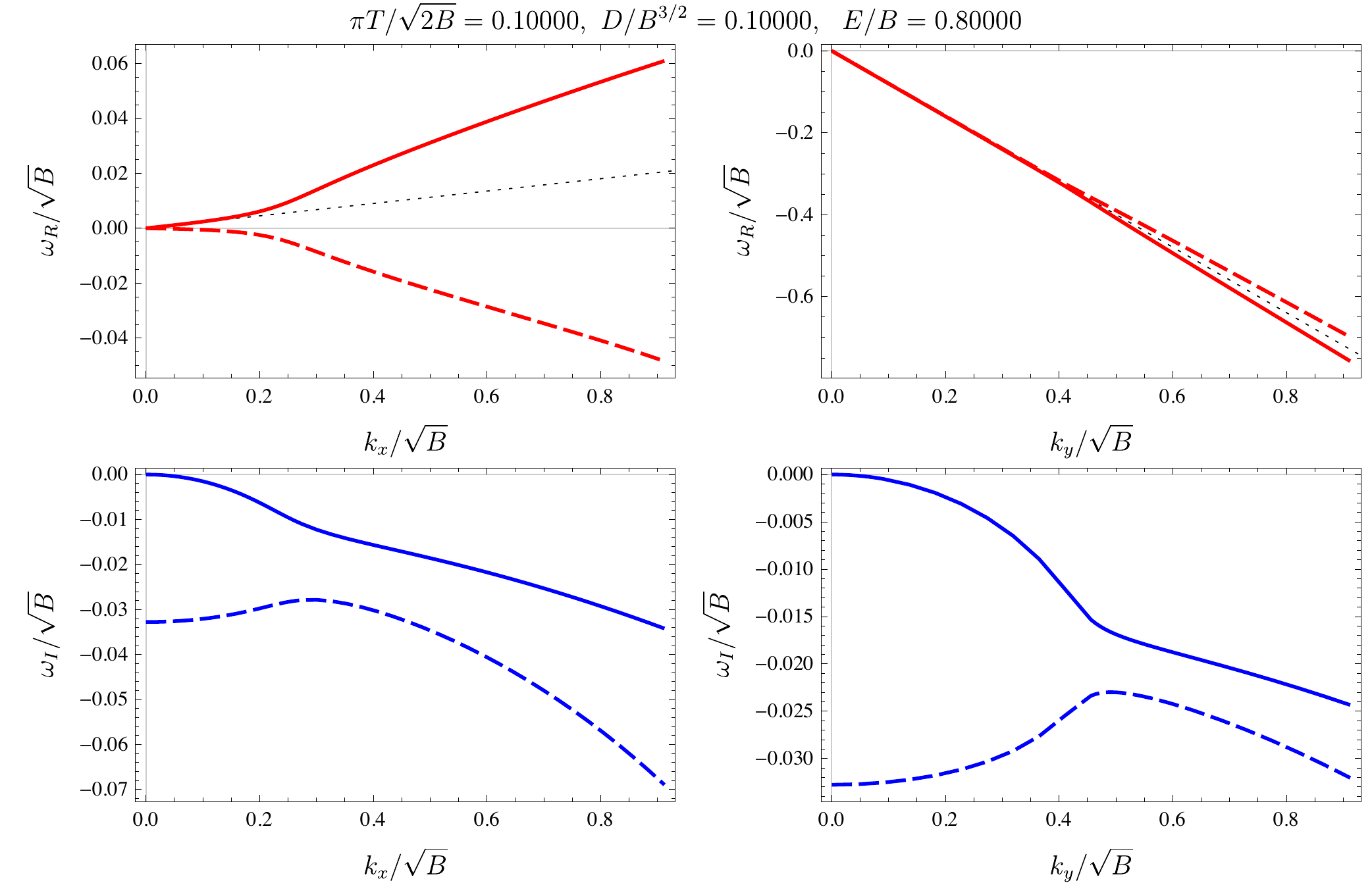}
	\caption{The behavior of the real part\,(upper panel) and the imaginary part\,(lower panel) of $\omega/\sqrt{B}$ as a function of $k_{x}/\sqrt{B}$\,(left panel) and $k_{y}/\sqrt{B}$\,(right panel) in the non-equilibrium state. The solid lines and the dashed lines denote the gapless modes and the gapped modes, respectively.
	}
	\label{fig:dispNESS2}
\end{figure}

We also show the dispersion relation for another values of the background parameters in Fig.\,\ref{fig:dispNESS2}.
In this choice of the parameters, we find that the gapless mode and the gapped mode with respect to $k_{x}$ have a positive linear dispersion in the real part of frequency. Therefore, we can also write the dispersion relation of these modes with respect to $k_{x}$ as the following form for small momenta: $\omega=\beta_{x}k_{x}+\order{k_x^2}$ and $\omega=-i/\tau + \tilde{\beta_{x}}k_{x}+\order{k_x^2}$ with a real-valued constant $\beta_{x}, \tilde{\beta_{x}}$, and $\tau$.

In the following section, we study the linear dispersion at small momentum for the gapless mode in the analytical approach.

\subsection{Linear dispersion}\label{sec:linear}
To understand these behaviors, we analytically study the linear dispersion in the limit of small frequency and momentum, namely hydrodynamic approximation. In the hydrodynamic approximation, we expand the fluctuations in a series in small $\omega$ and $k$:
\begin{eqnarray}
	{\cal W}(u) &=& {\cal W}^{(0)}(u)+{\cal W}^{(1)}(u)+{\cal W}^{(2)}(u)+ \cdots, \\
	{\cal A}_{z}(u) &=& {\cal A}_{z}^{(0)}(u)+{\cal A}_{z}^{(1)}(u)+{\cal A}_{z}^{(2)}(u)+ \cdots,
\end{eqnarray}
where the indices denote the order of $\omega$ and $k$. Substituting these expansions into the equations of motion for the fluctuations (\ref{eq: linearized eom E}), we can solve them order by order.
At the zeroth order, the equations are given by
\begin{equation}
	\partial_{u} \left[
		{\cal F}_{w}
		\gamma^{uu} \partial_u \frac{\mathcal{W}^{(0)}(u)}{y(u)}
	\right] =0,~~~
	\partial_{u} \left[
		{\cal F}_{\cal A}
		\gamma^{uu} \partial_{u} \mathcal{A}^{(0)}(u)
	\right] = 0.
\end{equation}
The solution for the former equation is written as
\begin{equation}
	\mathcal{W}^{(0)}(u)
	=
	y(u)\left[
		\tilde{C}^{(0)} \int_{u_*}^{u} \frac{\dd u'}{{\cal F}_{w} \gamma^{uu}}
		+ C^{(0)}
	\right],
\end{equation}
where \(\tilde{C}^{(0)}\) and \(C^{(0)}\) are an integration constant.
Since the first integral diverges at \(u=u_*\), the first term must vanish for the regularity condition.
So we set \(\tilde{C}^{(0)}=0\).
Another constant can be determined \(C^{(0)}=\mathcal{W}(u_*)/y(u_*)\) by imposing \(\mathcal{W}^{(0)}(u_*) = \mathcal{W}(u_*)\).
As those, we obtain \({\cal A}^{(0)}_z(u) = {\cal A}_z(u_*)\) at the zeroth order.
%\({\cal W}(u) \propto y(u)\) in the limit of \(\omega = k_x = k_y = 0\) explicitly means the existence of the NG mode in spontaneously \(\chi\)SB phase.
We obtain the solutions up to the first order in \(\omega\) and \(k\) as
\begin{equation}
\begin{bmatrix}
	\mathcal{W}(u) \\
	\mathcal{A}_z(u)
\end{bmatrix}
=
\begin{bmatrix}
	\frac{y(u)}{y(u_*)} + \Pi^{(1)}_{ww}(u)&
	\Pi^{(1)}_{w\mathcal{A}}(u) \\
	\Pi^{(1)}_{\mathcal{A} w}(u)&
	1 + \Pi^{(1)}_{\mathcal{A}\mathcal{A}}(u)
\end{bmatrix}
\begin{bmatrix}
	\mathcal{W}(u_*) \\
	\mathcal{A}_z(u_*)
\end{bmatrix}+ \cdots,
\label{eq: hydro approx solution}
\end{equation}
where
\begin{subequations}
\begin{alignat}{3}
	\Pi^{(1)}_{ww}(u)=& ~&
	ik_{\alpha} \frac{y(u)}{y(u_*)}&
	\int_{u_*}^{u} \dd u' \frac{{\cal F}_{w} \gamma^{u\alpha} - {\cal F}_{w} \gamma^{u\alpha}(u_*)}{{\cal F}_{w} \gamma^{uu}},\\
	\Pi^{(1)}_{w\mathcal{A}}(u)=&~&
	i(k_{y} E + \omega B)y(u)&
	\int_{u_*}^{u} \dd u' \frac{\cos^4 \theta - \cos^4 \theta(u_*)}{{\cal F}_{w}\gamma^{uu}},\\
	\Pi^{(1)}_{\mathcal{A}\mathcal{A}}(u)=&~&
	ik_{\alpha}&
	\int_{u_*}^{u} \dd u' \frac{{\cal F}_{\cal A}\gamma^{u\alpha} - {\cal F}_{\cal A} \gamma^{u\alpha}(u_*)}{{\cal F}_{\cal A} \gamma^{uu}},\\
	\Pi^{(1)}_{\mathcal{A} w}(u)=&~&
	\frac{i(k_{y} E + \omega B)}{y(u_*)}&
	\int_{u_*}^{u} \dd u' \frac{\cos^4 \theta - \cos^4 \theta(u_*)}{{\cal F}_{\cal A}\gamma^{uu}}.
\end{alignat}
\end{subequations}
The dispersion relation is given by an equation that the determinant of the matrix in (\ref{eq: hydro approx solution}) equals to zero at the boundary \cite{Kaminski2009d}.
Up to the linear order, this condition yields
\begin{equation}
	0 = \lim_{u\to 0}\left[
		\frac{y(u)}{y(u_*)}\left(
			1 + \Pi^{(1)}_{\mathcal{A}\mathcal{A}}(u)
		\right)
		+ \Pi^{(1)}_{ww}(u) + \cdots
	\right].
\end{equation}
We neglected a term \(-\Pi_{w\mathcal{A}}^{(1)}\Pi_{\mathcal{A}w}^{(1)}\) since it is contributed from the second-order of \(\omega\) and \(k\).
In massless cases, the first term vanishes because the embedding function is expanded as \(y(u) = \theta_2 u^2 + \order{u^3}\).
The second term gives the following series expansion in the vicinity of \(u=0\):
\begin{equation}
	\Pi^{(1)}_{ww}(u) = i k_{\alpha} \frac{{\cal F}_{w} \gamma^{u\alpha}(u_*)}{2\theta_2 y(u_*)} + \order{u^2}.
\end{equation}
Therefore, the dispersion relation in massless cases is written as
\(
		0 = i k_{\alpha} \gamma^{u\alpha}(u_{*}) + \cdots
\).
As a result, the linear dispersion for $k$ is given by
 \begin{equation}
 	\omega=-\frac{\gamma^{ux}}{\gamma^{ut}}k_{x}-\frac{\gamma^{uy}}{\gamma^{ut}}k_{y} + \cdots
 	\label{eq:linear}
\end{equation}
where dots denote the higher order in $k_{x}$ and $k_{y}$. Note that all the open string metric are evaluated at $u=u_{*}$.\footnote{
	%Notice that the linear dispersion we find can not be written as \(\omega = \pm v k\).
	In the present case, the modes are no longer symmetric with respect to flipping the sign of the real-part of \(\omega\).
	%This lack of time-reversal symmetry of the modes may be corresponding to the present system is far from equilibrium.
}
%The coefficients of $k_{x}$ and $k_{y}$ give the velocity with respect to each direction.
We write the linear coefficients as \(\beta_{x}\) and \(\beta_{y}\) for each direction.
By using the components of the inverse of the open string metric in Appendix \ref{apdx: osm}, we obtain
\begin{equation}
	\beta_{x} \equiv -\frac{\gamma^{ux}}{\gamma^{ut}} =
	\frac{
			B g_{tt} {a_y'}
			+E g_{xx} {a_t'}
		}{E g_{xx} {a_x'}}(u_*),~~~
	\beta_{y} \equiv -\frac{\gamma^{uy}}{\gamma^{ut}} =
	 \frac{B}{E}\frac{g_{tt}}{g_{xx}}(u_*).
\end{equation}
Substituting the metric (\ref{eq:metric}) and the solutions for the gauge fields (\ref{eq:gauge_sol}), we find
\begin{equation}
	\beta_{x} = \frac{B J_y + D E}{E J_x}f(u_*)
	=-\frac{J_y}{J_x}\left(\frac{E}{B}+\frac{J_y}{D}\right), ~~~
	\beta_{y} = -\frac{B}{E}f(u_{*})
	=\frac{J_y}{D}.
	\label{eq:coefs}
\end{equation}
We used \(f(u_*)=-\frac{E}{B}\frac{J_y}{D}\) from \(a(u_*)=0\).
%\(u_*\) is given by (\ref{eq:zstar}). $f(u_{*})$ is written as the function of $e$ and $b$:
%\begin{equation}
%	f(u_{*})=1-\frac{1}{b^{2}}\left(\frac{b^{2}-e^{2}-1}{2}+\sqrt{\left(\frac{b^{2}-e^{2}-1}{2}\right)^{2}+b^{2}} \right).
%\end{equation}
Here, we emphasize that the coefficients of $k_{x}$ and $k_{y}$ are written as the analytic functions of the quantities in the dual field theory.
These analytic results agree with the numerical results. Namely, the coefficients of $k_{x}$ and $k_{y}$ are consistent with numerical plot of the dispersion relation as denoted by the black dotted line in Fig.\,\ref{fig:dispNESS} and Fig\,\ref{fig:dispNESS2}.
Note that the mode obtained here is corresponding to the gapless mode.
On the other hand, the gapped mode and the splitting behavior in Fig.\,\ref{fig:disp1} can be obtained from the analysis in the second order of \(\omega\) and \(k\).
We can also compute the second order contributions of the dispersion relation, such as \(D_{\perp}\) in (\ref{eq:telegraph}).
We leave it as a future work.

Let us discuss the result for \(\beta_y\).
From Fig.\,\ref{fig:dispNESS} and Fig.\,\ref{fig:dispNESS2}, both of the gapless mode and the gapped mode for \(y\)-direction have same linear coefficient for \(\omega_R\) against \(k_y\).
This implies these modes are drifted by the velocity of \(\beta_y\).%
\footnote{
	The authors thank Referee and N.~Tanahashi for giving a comment on this point.
}
By considering the Lorentz boost with \(\beta_y\), we can change the coordinate system to the rest frame of these modes.
\footnote{%
	We show the dispersion relation in the rest frame of the NG modes in Appendix \ref{apdx:boost}.
}%
Then, we have the relationships:
\begin{equation}
	\hat{D} = \gamma_y (D - \beta_y J_y),~~
	\hat{J}_x= J_x,~~
	\hat{J}_y=\gamma_y (J_y - \beta_y D),~~
	\hat{J}_z= J_z,
\end{equation}
where \(\hat{J}^{\mu} = (\hat{D}, \hat{J}_x, \hat{J}_y, \hat{J}_z)\) is the four-current density in the rest frame of the modes.
\(\gamma_y\) is a Lorentz factor given by \(1/\sqrt{1-\beta_y^2}\).
Remarkably, we find \(\hat{J}_y=0\) by using (\ref{eq:coefs}) in this frame.
This means that the NG modes are drifted by the Hall current.

The expression for \(\beta_y\) is also written as
\begin{equation}
	\beta_y = - \frac{b}{e}\left(
		1 -
		\frac{b^2-e^2-1}{2 b^2}
		-\frac{1}{b^2}\sqrt{
			\left(\frac{b^2-e^2-1}{2}\right)^2 + b^2
		}
	\right),
\end{equation}
where \(e=E/\pi^2 T^2\) and \(b= B/\pi^2 T^2\).
Let us see the property of \(\beta_{y}\) as a function of \(e\) and \(b\).
This is odd function in both of \(e\) and \(b\).
For fixed \(b>0\), the maximum point of \(\beta_{y}(b,e)\) is given by \(e_{\text{max}} = -\sqrt{1+b^2}\).
Then, the maximum value of \( \beta_{y} \) is \(\beta_{y}(b,e_{\text{max}}) = (\sqrt{b^2+1}-1)/b\).
From this result, we can conclude that \(-1 < \beta_{y}<1\) for any \(e\) and \(b\).
On the other hand, \(\beta_{x}\) depends on \(J_x\) and \( J_y\) determined by the numerical computation for fixed mass.
We can not analytically check that \(\beta_{x}\) is always less than the speed of light.

One can check that the expressions in (\ref{eq:coefs}) vanish in the limit of $E \rightarrow 0$.%
\footnote{If \(f(u_*)\) is expanded in powers of \(E\), the leading order is \(E^2\).}
Thus, we find that there is no contribution of a linear order in the equilibrium background. This is also consistent with the dispersion relation in Fig.\,\ref{fig:disp1}.

\begin{figure}[tbp]
	\centering
    \includegraphics[width=14.0cm]{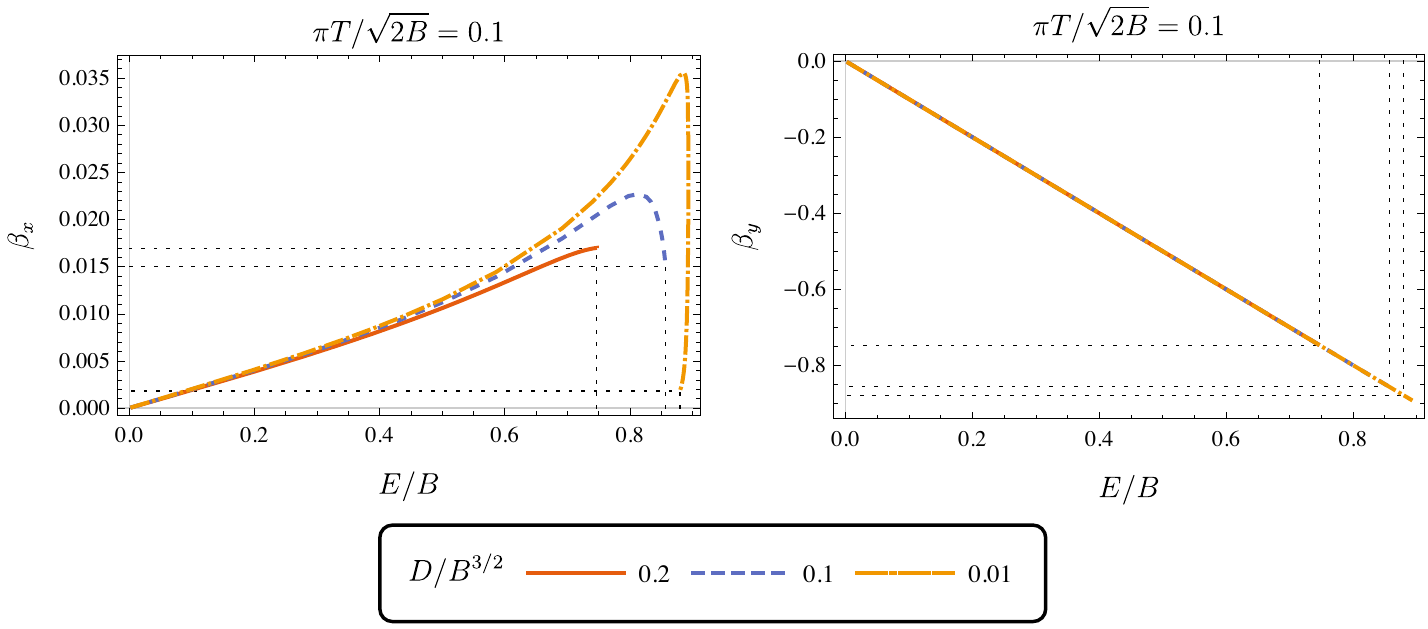}
	\caption{The behavior of $\beta_{x}$\,(left) and $\beta_{y}$\,(right) as functions of $E/B$ for several values of $D/B^{3/2}$ with $\pi T/ \sqrt{2B}$.
	The dotted lines show the endpoints of the \(\chi\)SB branch for each case.
	}
	\label{fig:betaVSE}
\end{figure}
Here, let us see the behaviors of $\beta_{x}$ and $\beta_{y}$ with respect to several parameters from (\ref{eq:coefs}). In Fig.\,\ref{fig:betaVSE}, we show $\beta_{x}$ and $\beta_{y}$ as functions of $E/B$ with temperature fixed. In the behavior of $\beta_{x}$, it is found that linear dispersion with respect to $k_{x}$ is small compared to $\beta_{y}$ but becomes finite. 
%In fact, we confirmed that the dispersion relation with respect to $k_{x}$ also has a linear dispersion as shown in Fig.\,\ref{fig:dispNESS2}.
As can be seen from the right panel of Fig.\,\ref{fig:betaVSE}, the behavior of $\beta_{y}$ with respect to $E/B$ are the same as that for different values of $D/B^{3/2}$.
This is because \(\beta_y\) is independent of $D$.
\begin{figure}[tbp]
	\centering
    \includegraphics[width=14cm]{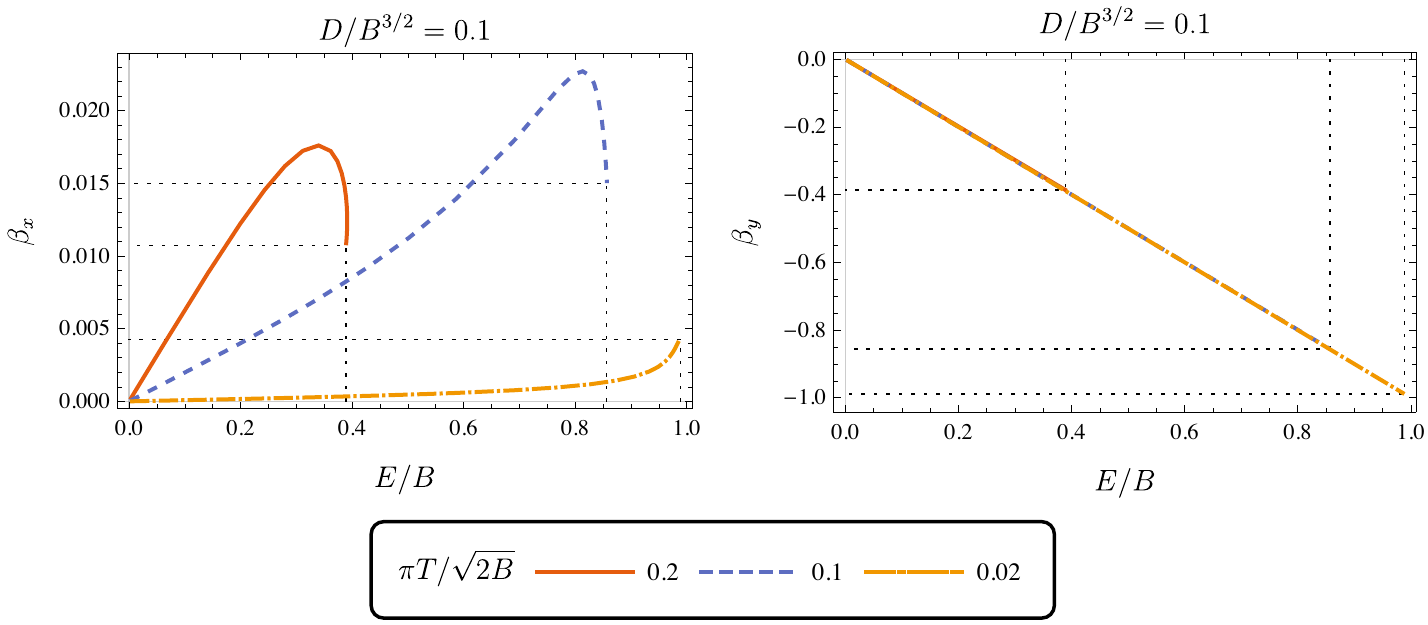}
	\caption{The behavior of $\beta_{x}$\,(left) and $\beta_{y}$\,(right) as functions of $E/B$ for several values of $\pi T/ \sqrt{2B}$ with $D/B^{3/2}$.
	The dotted lines show the endpoints of the \(\chi\)SB branch for each case.
	}
	\label{fig:betaVSE2}
\end{figure}
We also show $\beta_{x}$ and $\beta_{y}$ as functions of $E/B$ with charge density fixed in Fig.\,\ref{fig:betaVSE2}.
We also find that $\beta_{x}$ has small value for several values of $\pi T/ \sqrt{2B}$.
While $\beta_{y}$ depends on $\pi T/ \sqrt{2B}$, the behavior for each temperature is almost same each other.
This is because \(E/B\) and \(\pi T/\sqrt{2B}\) are small for each \(\chi\)SB solution.
For small \(E/B\), \(\beta_{y}\) is given by \(\beta_{y}\approx -\frac{1}{1+4\tilde{w}_H^{4}}\frac{E}{B} \) where \(\tilde{w}_H=\pi T/\sqrt{2B}\).
The slope of \(\beta_y\) with respect to \(E/B\) clearly depends on \(\tilde{w}_H\) but this is almost \(1\) for small \(\tilde{w}_H\).
%This can be understood from the following consideration.
%The $\beta_{y}$ can be explicitly written as
%\begin{equation}
%   \beta_{y}=\frac{1+\tilde{E}^{2}+4\tilde{w}_{H}^{4}-\sqrt{\tilde{E}^{4}+2\tilde{E}^{2}(4\tilde{w}_{H}^{4}-1)+(4\tilde{w}_{H}^{4}+1)^2}}{2\tilde{E}},
%\end{equation}
%where $\tilde{E}=E/B$ and $\tilde{w}_{H}=\pi T/ \sqrt{2B}$ are parameters scaled by $B$. If one expands this in $\tilde{E}$, the $\beta_{y}$ is given by
%\begin{equation}
%    \beta_{y}=-\frac{1}{1+4\tilde{w}_{H}^{4}}\tilde{E} + {\cal O}(\tilde{E}^{2}). 
%\end{equation}
%The slope of $\beta_{y}$ with respect to $\tilde{E}$ clearly depends on $\tilde{w}_{H}$. However, $\tilde{w}_{H}<1$ in the $\chi$SB phase and the coefficient of $\tilde{E}$ is almost $1$ for each %temperature. Also, the phase transition occurs for large $\tilde{E}$ and/or large $\tilde{w}_{H}$. %Therefore, the behavior of $\beta_{y}$ seems to be almost same for each temperature.

\section{Conclusion and Discussion}  \label{sec5}
In this paper, we study the dispersion relation of the NG mode which appears in the dissipative system. The system is holographically realized by using the D3/D7 model in the presence of a charge density and an external magnetic field. In the dual field theory, the system is in equilibrium, but the fluctuations dissipate into the heat bath. In this system, there is a phase transition associated with the $U(1)$ chiral symmetry. We analyze the dispersion relation of the NG mode associated with the spontaneous symmetry breaking of the $U(1)$ chiral symmetry by studying the behaviors of QNMs in the dual gravity. For small momentum, we find that the NG mode behaves as a diffusive mode. This result agrees with recent works \,\cite{Minami2018,Hongo:2019qhi} in which the dispersion relation of the NG mode in a dissipative system was studied in terms of the effective field theory. Following the Ref.\,\cite{Minami2018,Hongo:2019qhi}, the dispersion relation of the NG mode in our setup is classified into the type-A NG mode. If we increase momentum, these two modes are no longer purely imaginary and the real parts of them become finite. In other words, the diffusive regime is changed into the reactive regime. This specific behavior has been known in the dispersion relation observed in the dissipative system\,\cite{Baggioli2019}. For large momentum, however, we find that these modes turn into purely imaginary modes again. The non-trivial behavior for large momentum seems to be related to the presence of the charge density since it will vanish in the limit of zero charge density. We emphasize that these behaviors can be observed since our analysis contains the higher order contribution of momentum to the dispersion relation.

Moreover, we also study the dispersion relation of the NG mode in a NESS system. We also apply an external electric field which is perpendicular to the magnetic field. As a result, the system has a constant electric current along the electric field.
In addition, there is also constant electric current along to the direction which is perpendicular to both of the electric field and magnetic field by Hall effect.
In this NESS system, we investigate the dispersion relation of the NG mode associated with the spontaneous $U(1)$ chiral symmetry breaking.
We find that the dispersion relation is similar to that in the equilibrium background. That is, the gapless mode shows the diffusive behavior for small momentum even in the NESS system. Analyzing the linear dispersion, the velocity with respect to $k_{y}$ becomes large, whereas that with respect to $k_{x}$ is very close to zero.
%However, the dispersion relation with respect to $k_{y}/\sqrt{B}$ shows that these modes represent a linear dispersion at small momentum.
This behavior also agrees with the propagating mode for the type-A NG mode as discussed in Ref.\,\cite{Minami2018,Hongo:2019qhi}. Moreover, the hydrodynamic approximation shows that the coefficients of $k_{x}$ and $k_{y}$ in the linear dispersion are written as the analytic functions of quantities in the dual field theory.
As shown in the hydrodynamics approximation, the linear dispersion is arising from the off-diagonal components of the open string metric.
In the gravity side, the existence of the off-diagonal components are corresponding to the system driven far from equilibrium.
%On the other hand, if we analyze the dispersion relation of Minkowski embedding in which the chiral symmetry is broken without dissipation in the presence of both the magnetic field and the electric field, we obtain the propagating NG mode with linear dispersion. This implies that a linear dispersion itself stems from an anisotropy of the system rather than a NESS regime. It is not clear whether an anisotropy of the system is necessary to obtain a linear dispersion.
Therefore, we expect that the linear dispersion observed in our system is characteristic behavior in non-equilibrium steady state.

Before closing this section, we have some remarks. In this paper, we consider the momentum components of $(k_{x},k_{y},0)$ for simplicity. However, it would be interesting to study the $k_{z}$-dependence of the dispersion relation. Considering the fluctuations with finite $k_{z}$ in the D3/D7 model with a finite charge density and a magnetic field, the chiral helix phase was found in the chiral symmetry restored phase\,\cite{Kharzeev2011}. Since we expect that this can be seen even in the chiral symmetry breaking phase, it would be interesting to analyze the corresponding QNMs with finite $k_{z}$. 
In this paper, we focus only on the NG modes associated with spontaneous symmetry breaking of $U(1)$ chiral symmetry. On the other hand, the explicit symmetry breaking and the pseudo-NG modes have been studied in this model\,\cite{Filev2009}. It would be interesting to study the interplay of explict and spontaneous symmetry breaking in our setup as discussed in\,\cite{Baggioli2019b}.
Also, we expect that the linear dispersion of Eq.\,(\ref{eq:linear}) which depends only on quantities of the dual field theory could be experimentally observed. More precisely, our results suggest that the linear dispersion of NG mode associated with the $U(1)$ chiral symmetry breaking should be determined by the NESS background in which there is a charge density in the presence of an external electric field and magnetic field.

\section*{Acknowledgement}
The authors are grateful to Y.~Hidaka, M.~Hongo, H.~Hoshino, S~ Nakamura, S.~Kinoshita, N.~Tanahashi, and R.~Yoshii for helpful discussions and comments.
The authors thank M.~Baggioli and B.~Gout\'eraux for providing comments.
M.~M.~is supported by National Natural Science Foundation of China Grants No.~12047538.

\begin{appendices}
\section{Open string metric}\label{apdx: osm}
As we discussed in section \ref{sec4}, we assume that the fields components have the following form:
\begin{equation}
	\theta=\theta(u), \hspace{1em} \psi=0, \hspace{1em} A_{t}=a_{t}(u),\hspace{1em} A_{x}=-Et+a_{x}(u),\hspace{1em} A_{y}=Bx + a_{y}(u).
\end{equation}
The open string metric is defined by $\gamma_{ab}=g_{ab}-(Fg^{-1}F)_{ab}$, where $g_{ab}$ is the induced metric on the D7-brane.
The equations of motion on the worldvolume depend on the inverse of the open string metric:\,\(\gamma^{ab}\).
Here, we write the non-zero components of the inverse of the open string metric in our setup.
The diagonal components are written as
\begin{subequations}
\begin{align}
	\gamma^{tt}=&
	\frac{
		-g_{xx} \left(
			g_{xx} ( {a_x'}^2+{a_y'}^2 )
			+g_{uu} (g_{xx}^2+B^2)
		\right)
	}{R(u)},\\
	\gamma^{xx}=&\frac{
		g_{xx} \left(
			-g_{xx} {a_t'}^2
			- g_{tt} ({a_y'}^2+g_{uu} g_{xx})
 		\right)
 	}{R(u)},\\
 	\gamma^{yy}=&\frac{
 		g_{xx} \left(
 			-g_{xx} {a_t'}^2
 			-g_{tt} {a_x'}^2
 			+g_{uu} (-g_{tt} g_{xx}-E^2)
 		\right)
 	}{R(u)},\\
 	\gamma^{zz}=& \frac{1}{g_{xx}},\\
 	\gamma^{uu}=&
 	\frac{\xi(u)}{R(u)},
\end{align}
\end{subequations}
where \(\xi(u)\) is defined by (\ref{eq:xi}) and \(R(u)\) is defined by
\begin{equation}
\begin{aligned}
	R(u)=&
	- \frac{\det( g_{ab} + F_{ab})}{\cos^6 \theta}\\
	=&
	g_{uu} \xi(u)
	+g_{xx} \left(
		2 B E {a_t'} {a_y'}
		-{a_t'}^2 ( g_{xx}^2+B^2)
		-g_{tt} g_{xx} ({a_x'}^2+{a_y'}^2)
		-E^2 {a_y'}^2
	\right).
\end{aligned}
\end{equation}
The off-diagonal components are written as
\begin{subequations}
\begin{align}
	\gamma^{tx} =&\frac{ g_{xx}^2 {a_t'} {a_x'} }{R(u)},\\
	\gamma^{ty} =&
	\frac{
		g_{xx} \left(g_{xx} {a_t'} {a_y'}+B
		   E g_{uu}\right)
	}{R(u)},\\
	\gamma^{xy} =& \frac{
		g_{tt} g_{xx} {a_x'}{a_y'}
	}{R(u)}
\end{align}
\end{subequations}
and
\begin{subequations}
\begin{align}
	\gamma^{tu}=&
	\frac{-E g_{xx}^2 {a_x'}}{R(u)},\\
	\gamma^{xu}=&
	\frac{
		g_{xx} \left(E g_{xx} {a_t'}+B
		   g_{tt} {a_y'}\right)
	}{R(u)},\\
	\gamma^{yu}=&
	\frac{
		-B g_{tt} g_{xx} {a_x'}
	}{R(u)}.
\end{align}
\end{subequations}
Since the open string metric is symmetric, another off-diagonal components are also given by the above expressions.
The other components are zero.

%\begin{equation}
%	ds^{2}_{\rm osm}=\gamma_{tt} d\tilde{t}^{2}+2\gamma_{tx} d\tilde{t}d\tilde{x}+\gamma_{xx}d\tilde{x}^{2}+(\gamma_{uu}-X)du^{2}+\cdots,
%\end{equation}
%where
%\begin{eqnarray}
%	&&d\tilde{t}\equiv dt+\frac{\gamma_{tx}\gamma_{xu}-\gamma_{tu}\gamma_{xx}}{2(\gamma_{tx}^{2}-\gamma_{tt}\gamma_{xx})}du, \hspace{1em} d\tilde{x}\equiv dx+\frac{\gamma_{tx}\gamma_{tu}-\gamma_{tt}\gamma_{xu}}{2(\gamma_{tx}^{2}-\gamma_{tt}\gamma_{xx})}du, \\
%	&&X\equiv \frac{2\gamma_{tu}\gamma_{tx}\gamma_{xu}-\gamma_{tt}\gamma_{xu}^{2}-\gamma_{xx}\gamma_{tu}^{2}}{4(\gamma_{tx}^{2}-\gamma_{tt}\gamma_{xx})}.
%\end{eqnarray}

\section{Dispersion relation of lower three modes}\label{apdx:disp}
We show more examples of the dispersion relation of the NG modes in a wide range of momentum in the equilibrium system.
We also find that the third mode connects to the dispersion relation of the NG mode in a few cases of the background configurations.

\begin{figure}[htbp]
	\centering
	\includegraphics[width=0.48\linewidth]{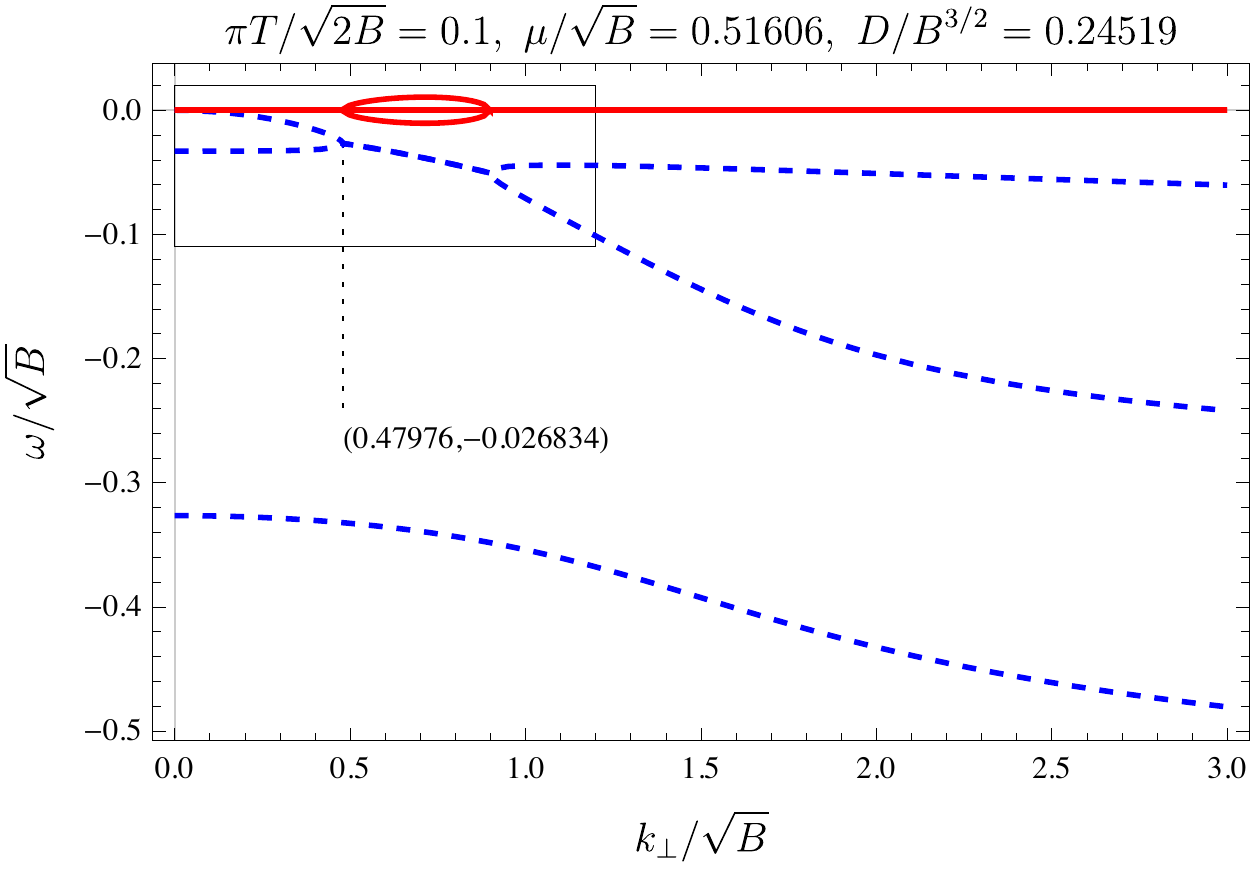}
	\includegraphics[width=0.48\linewidth]{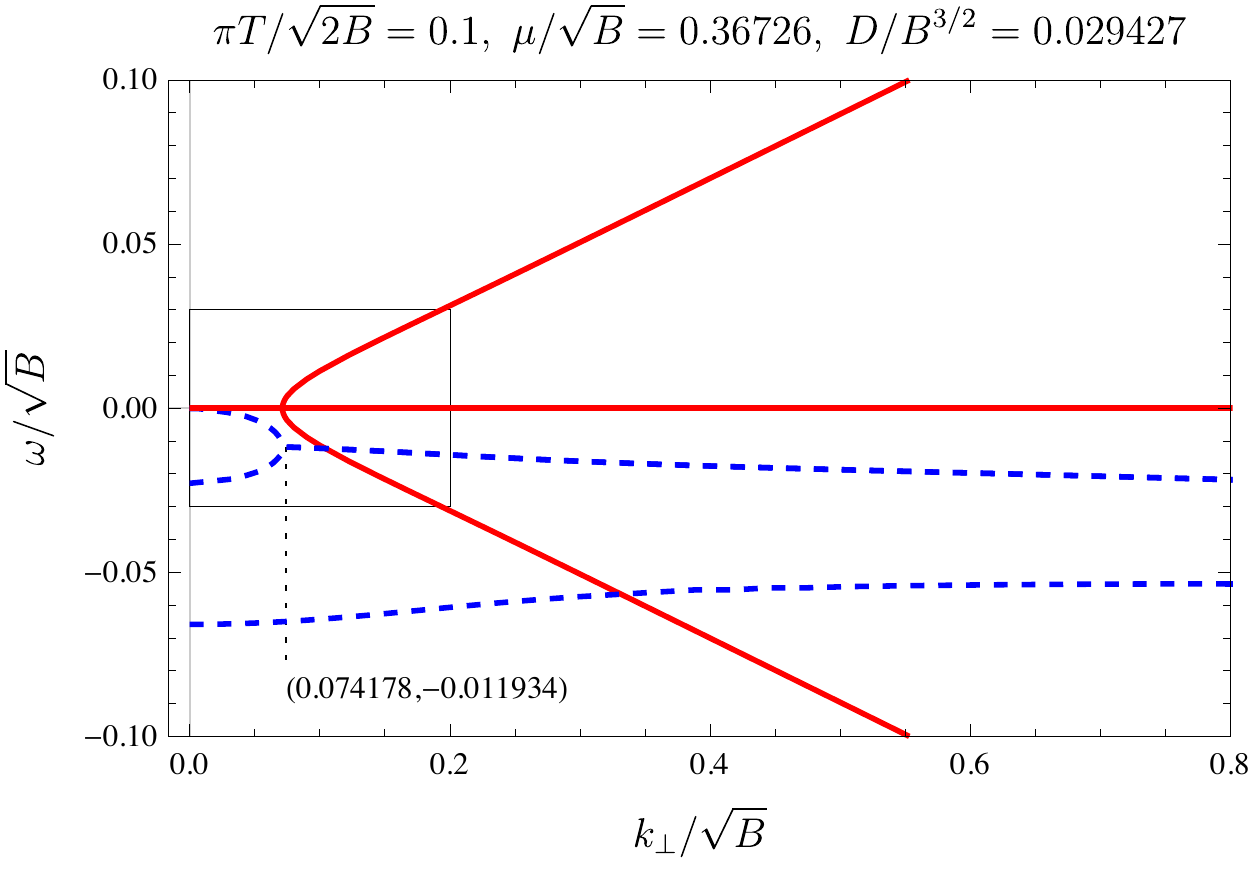}\\
	\includegraphics[width=0.48\linewidth]{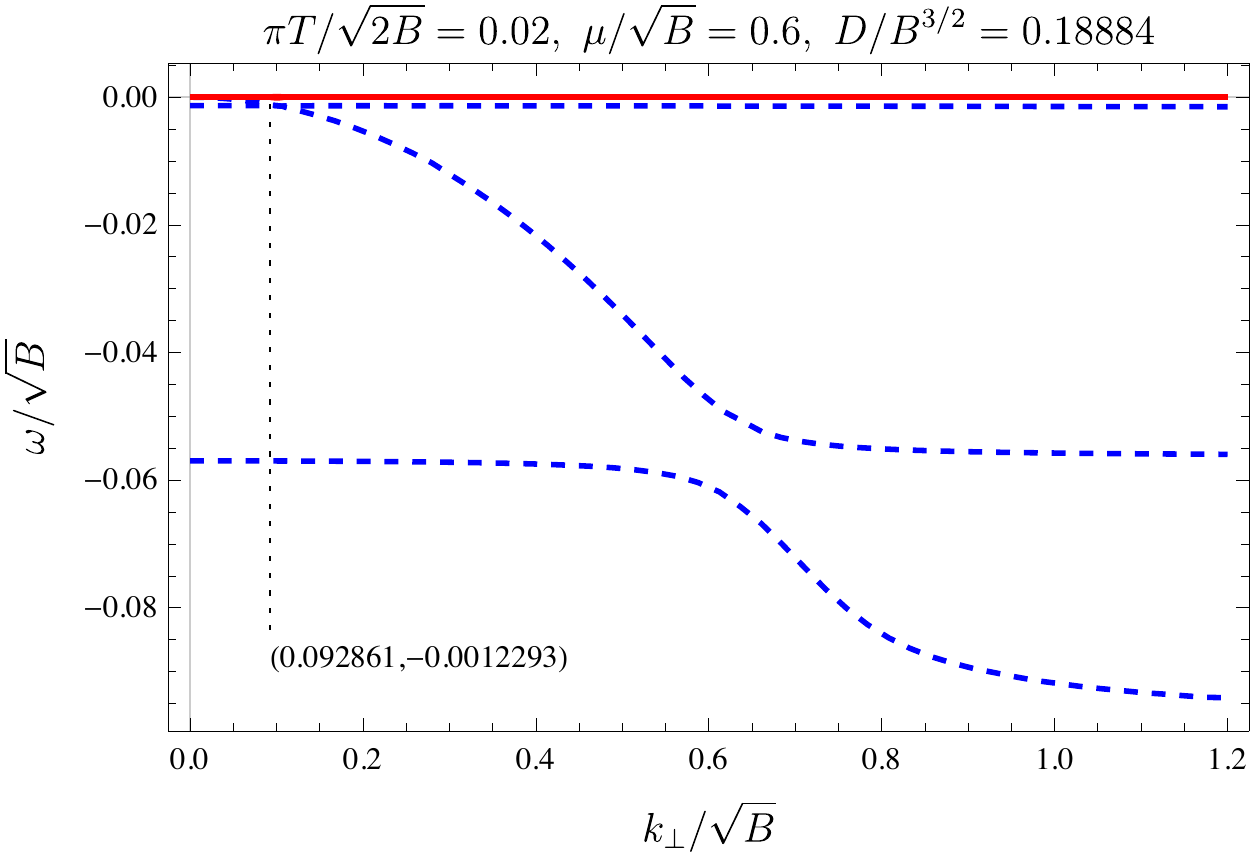}
	\includegraphics[width=0.48\linewidth]{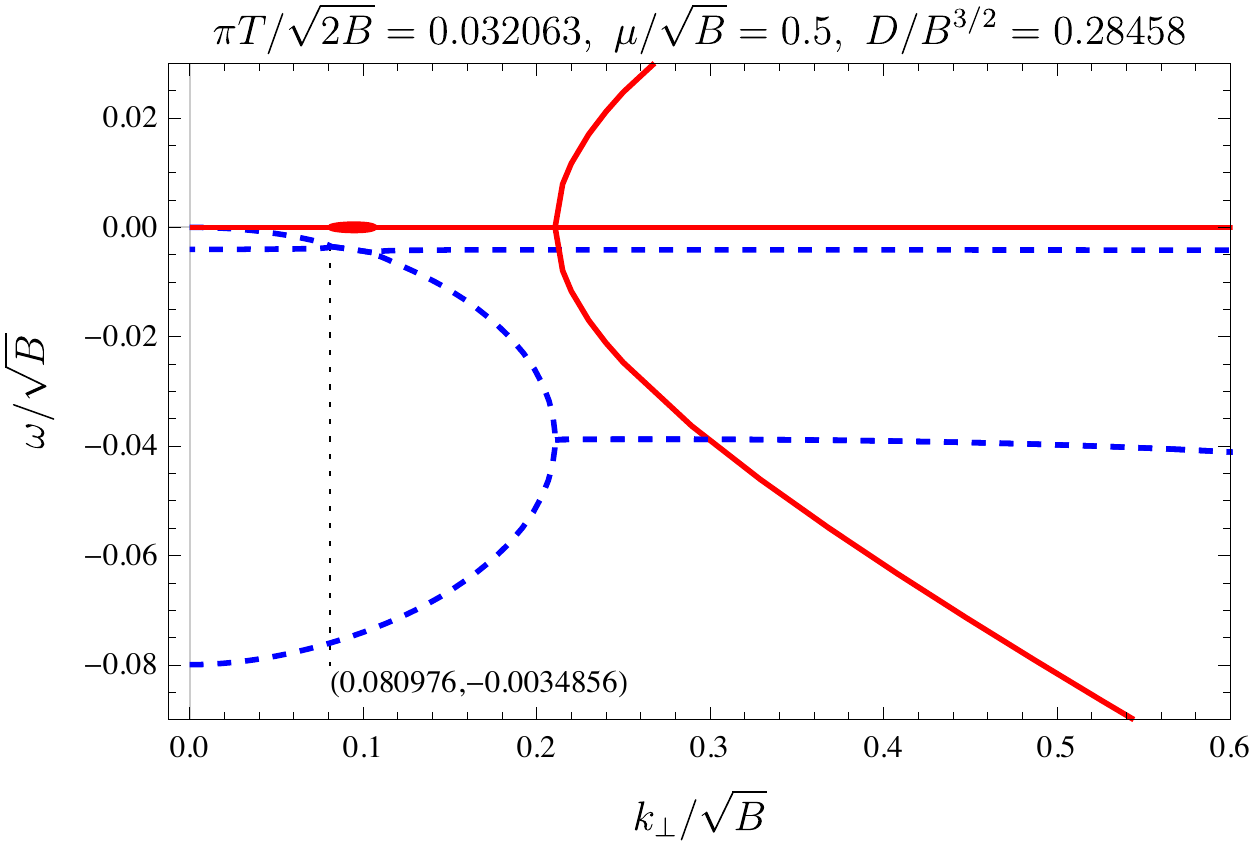}
	\caption{
	Dispersion relations of the lower three modes in \(\omega_{I}\).
	The solid lines and the dashed lines show the real part and the imaginary part of the modes, respectively.
	The dotted lines show the locations of the \(k\)-gap: \((k_{\text{gap}}/\sqrt{B}, \omega_I/\sqrt{B})\).
	The rectangle regions in the upper panels indicate the range of the plots in Fig.\,\ref{fig:disp1}.
	}
	\label{fig:disp_wide}
\end{figure}
The upper panels of Fig.\,\ref{fig:disp_wide} show the dispersion relations of the lower three modes in \(\omega_I\) for the background parameters already shown in Fig.\,\ref{fig:disp1}.
In these cases, the third purely imaginary mode does not connect to the dispresion relation of the NG modes.
The lower panels of Fig.\,\ref{fig:disp_wide} show the dispersion relations for other background parameters.
In these cases, the separation between \(k\)-gap and the re-merging point of \(\omega_R\) becomes very narrow.
Remarkably, the dispersion relation of the third mode is also connected to the NG modes in the lower-right panel of Fig.\,\ref{fig:disp_wide}.

We can also classify these dispersion relations by the behavior at enough large momentum.
In the left column of Fig.\,\ref{fig:disp_wide}, there are three purely imaginary modes at the right edge of \(k_{\perp}\)-axis.
In the right column of Fig.\,\ref{fig:disp_wide}, there are two modes with nonzero \(\omega_R\) and one purely imaginary mode at the right edge of \(k_{\perp}\).
However, we computed the dispersion relations of the three modes only for finite value of the momentum.
The full structure of the dispersion relations of the modes are not understood completely.
We leave it as an open question.

\section{Dispersion relation in the rest frame of the modes}\label{apdx:boost}
As discussed in section \ref{sec:linear}, the NG modes are drifted by the velocity \(\beta_y\) in the NESS.
By considering the Lorentz boost with \(\beta_y\) against \(y\)-direction, we obtain
\begin{equation}
	\hat{E} = \gamma_y (E-B\beta_y),~~
	\hat{B} = \gamma_y (B-E\beta_y)
\end{equation}
and
\begin{equation}
	\hat{\omega} = \gamma_y (\omega - \beta_y k_y),~~
	\hat{k}_{\hat{x}} = k_x,~~
	\hat{k}_{\hat{y}} = \gamma_y (\omega - \beta k_y),
\end{equation}
where \(\gamma_y=1/\sqrt{1-\beta_y^2}\) and hat\((\hat{~})\) denotes quantities in the boosted frame.
\begin{figure}[htbp]
	\centering
	\includegraphics[width=0.99\linewidth]{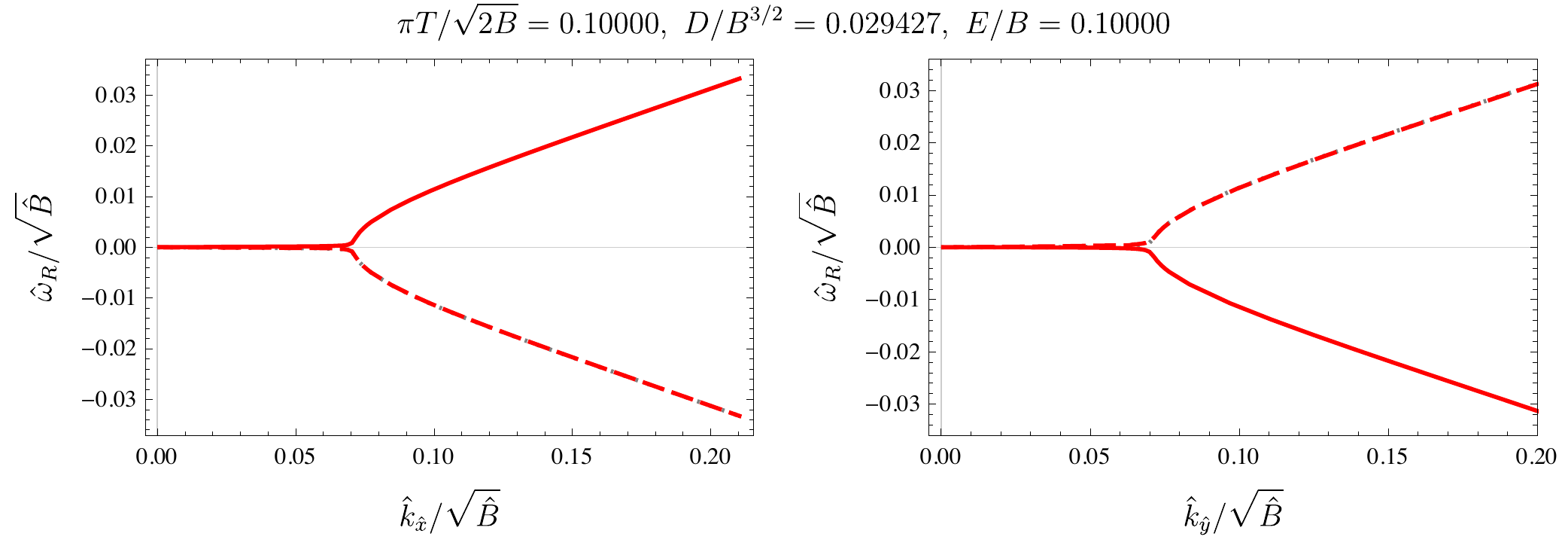}\\
	\includegraphics[width=0.99\linewidth]{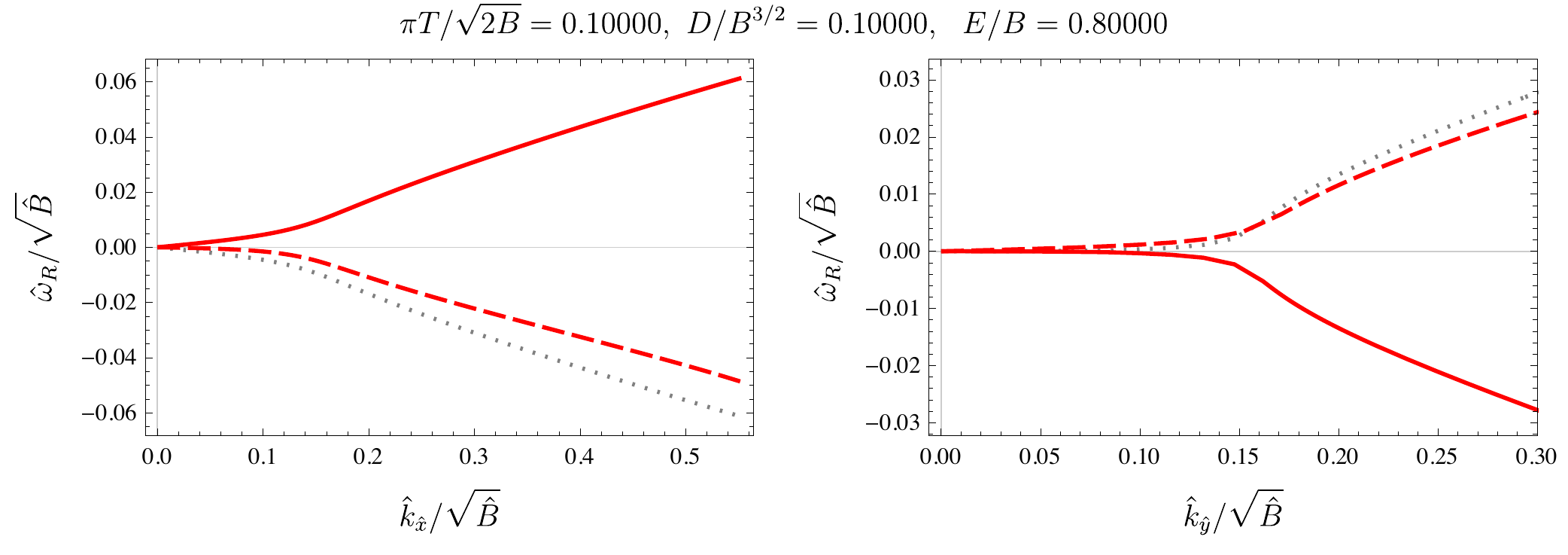}
	\caption{
		Dispersion relation of the NG modes in the rest frame of the modes.
		The background parameters are same as Fig.\,\ref{fig:dispNESS} (for upper panels) and Fig.\,\ref{fig:dispNESS2} (for lower panels).
		The solid line and the dashed line correspond to \(\hat{\omega}_R\) of the gapless mode and the gapped mode, respectively.
		The dotted line shows \(-\hat{\omega}_R\) of the gapless mode.
		In the upper panels, this is almost same as \(\hat{\omega}_R\) of the gapped mode.
	}
	\label{fig:dispNESS_boost}
\end{figure}

Figure \ref{fig:dispNESS_boost} shows the dispersion relations of the NG modes in the rest frame of the modes.
Here, \(\hat{\omega}=\hat{\omega}_R + i \hat{\omega}_I\) but we ignored the imaginary part of the modes for simplicity.
The linear coefficients of \(\hat{\omega}_R\) to \(\hat{k}_{\hat{y}}\) disappear due to the Lorentz boost with \(\beta_y\) for the gapless modes.
For \(E/B=0.1\) (upper panels of Fig.\,\ref{fig:dispNESS_boost}), the dispersion relation for \(\hat{x}\)- and \(\hat{y}\)-direction are almost same.
The flipping symmetry, \(\hat{\omega}_R\to -\hat{\omega}_R\), is also almost recovered.
In this case, the dispersion relation is similar to the one obtained in the equilibrium system.
For \(E/B=0.8\) (lower panels of Fig.\,\ref{fig:dispNESS_boost}), on the other hand, the dispersion relation is still anisotropic and not symmetric in \(\hat{\omega}_R \to -\hat{\omega}_R\).
We consider that these characteristic of the dispersion relation appears where \(D/B^{3/2}\) and \(E/B\) are large enough.

\end{appendices}

\end{document}